\documentclass[aps,onecolumn,letterpaper,oneside,preprint,tightenlines,draft,%
nobibnotes,nofootinbib,amsfonts,amssymb,amsmath,eqsecnum,showpacs,%
showkeys]{revtex4}
\usepackage[dvipdfm]{color,graphicx}
\newcommand{\cN}{\mathcal{N}}
\newcommand{\cP}{\mathcal{P}}
\newcommand{\cT}{\mathcal{T}}
\newcommand{\cV}{\mathcal{V}}
\newcommand{\cW}{\mathcal{W}}
\newcommand{\bH}{\boldsymbol{H}}
\newcommand{\bQ}{\boldsymbol{Q}}

\newcommand{\bmb}{\boldsymbol{b}}
\newcommand{\bmc}{\boldsymbol{c}}
\newcommand{\tH}{\tilde{H}}
\newcommand{\tP}{\tilde{P}}
\newcommand{\tR}{\tilde{R}}
\newcommand{\tcV}{\tilde{\cV}}
\newcommand{\htP}{\widehat{\tP}}
\newcommand{\bbC}{\mathbb{C}}
\newcommand{\rmi}{\mathrm{i}}%neIOPART
\newcommand{\rme}{\mathrm{e}}%neIOPART
\newcommand{\rmd}{\mathrm{d}}%neIOPART
\newcommand{\rmP}{\mathrm{P}}
\newcommand{\del}{\partial}

\newcommand{\lra}{\leftrightarrow}
\newcommand{\braket}[1]{\bigl\langle{#1}\bigr\rangle}

\begin{document}

%\markboth{Authors' Names}
%{Instructions for Typing Manuscripts (Paper's Title)}%WSPC

%\catchline{}{}{}{}{}%WSPC
%
%%%%%%%%%%%%%%%%%%%%%%%%%%%%%%%%%%%%%%%%%%%%%%%%%%%%%%%%%%%%%%%%%%%%

\title{Two-step Shape Invariance in the Framework of $\cN$-fold Supersymmetry}
%\article[Short title]{TYPE}{Full title}%IOPART
%ELSART,WSPC
%\author{},
%\ead{}%ELSART
%\address{}
%\author{Toshiaki Tanaka}
%\ead{toshiaki@post.kek.jp}%ELSART
%\address{Institute of Particle and Nuclear Studies,
% High Energy Accelerator Research Organization (KEK),
% 1-1 Oho, Tsukuba, Ibaraki 305-0801, Japan}
%REVTEX4
\author{Barnana Roy}
\email{barnana@isical.ac.in}
\affiliation{Physics and Applied Mathematics Unit, Indian Statistical
 Institute,\\ Kalkata 700108, India}
\author{Toshiaki Tanaka}
\email{toshiaki@post.kek.jp}
\affiliation{Institute of Particle and Nuclear Studies,
 High Energy Accelerator Research Organization (KEK),
 1-1 Oho, Tsukuba, Ibaraki 305-0801, Japan}
%IOPART
%\author{Author 1\dag, Author 2\ddag and Toshiaki Tanaka\S}
%\address{\dag Address 1}
%\address{\ddag Address 2}
%\address{\S Institute of Particle and Nuclear Studies,
% High Energy Accelerator Research Organization (KEK),
% 1-1 Oho, Tsukuba, Ibaraki 305-0801, Japan}
%\ead{toshiaki@post.kek.jp}
%\eads{\mailto{email 1}, \mailto{email 2},
% \mailto{toshiaki@post.kek.jp}}

%\date{\today}

\begin{abstract}

We extensively investigate two-step shape invariance in the framework of
$\cN$-fold supersymmetry. We first show that any two-step shape-invariant
system possesses type A $2$-fold supersymmetry with an intermediate Hamiltonian
and thus has second-order parasupersymmetry as well. Employing the general
form of type A $2$-fold supersymmetry, we systematically construct two-step
shape-invariant potentials. In addition to the well-known ordinary shape-invariant
potentials, we obtain several new and novel two-step shape-invariant ones which
are not ordinary shape invariant. Furthermore, some of the latter potentials are
conditionally two-step shape invariant and thus are conditionally solvable.
%\keywords{keyword1; keyword2; keyword3.}%WSPC
\end{abstract}

%\ccode{PACS Nos.: nn.nn.Xx; nn.nn.XX}%WSPC

%REVTEX4
%REVTEX4

%\begin{keyword}%ELSART
%  keyword 1\sep keyword 2\sep keyword 3
% \sep keyword 4\sep keyword 5\sep keyword 6
% \PACS nn.nn.Xx\sep nn.nn.Xx\sep nn.nn.Xn\sep nn.nn.Xx
%\end{keyword}%ELSART

%\pacs{nn.nn.Xx, nn.nn.Xx, nn.nn.Xx, nn.nn.Xx}%IOPART
%\submitto{\jpa}

\preprint{TH-1528}%REVTEX4

\maketitle

\section{Introduction}
\label{sec:intro}

In the formalism of supersymmetric quantum mechanics (SUSY QM) \cite{Wi82,CF83}
potentials with unbroken SUSY and shape invariance (SI) \cite{Ge83} can be
exactly solved by a well-known standard procedure \cite{IH51,GK85}. The potential
algebras of these systems have also been identified \cite{WAG89,GMS98,CDGPRS98,Ba98} 
providing an alternate method of finding exact solutions. Mainly three classes of SI have
been studied in the literature, namely, (1) translational class \cite{CGK87,DKS88,Le89,Ch91}
 - when the parameters differ only by a constant i.e. $b^{(1)}=b^{(0)}+\alpha$,
$\alpha$ being a constant, (2) scaling class \cite{BDGKPS93,KS93} - when the parameters
are related by a scaling factor i.e. $b^{(1)}=q b^{(0)}$ with $0<q<1$, and (3) exotic class
\cite{BDGKPS93} - when the parameters are related by other equalities such as
$b^{(1)}=q (b^{(0)})^{p}$ and $b^{(1)}=q b^{(0)}/(1+p b^{(0)})$.
It should be mentioned here that SI is neither a necessary nor a sufficient condition
for exact solvability. In fact, some exactly solvable potentials are shown not to be SI
\cite{KS93}. Recently a complete set of additive SI potentials have been
generated from an Euler equation \cite{BGM10}. The SI condition is also
studied in the context of fractional SUSY QM of order $k$ ($k=3,4, \ldots$) generalizing
the $Z_2$ grading of the relevant Hilbert space to a $Z_k$ grading \cite{DK02,CF03,%
DK04,DK06}.

The idea of SI can be extended to the more general concept of SI
in two and even multi-steps. While there are quite a number of works on
one-step (namely, ordinary) SI, the literature dealing with SI in two or more steps is
rather scanty. In \cite{BDGKPS93} the authors first introduced the concept of SI
in two steps to enlarge the class of exactly solvable Hamiltonians. The
two-step SI approach was utilized for dealing with SUSY QM problems
with spontaneously broken SUSY \cite{GMS01}. For a quantum mechanical system
with position-dependent mass the same approach was used to handle broken SUSY
problem \cite{MR09b}. Recently a class of solvable potentials of translational SI
in two steps were obtained \cite{Su08a,Su08b}. It was found that discontinuity at some
points was a characteristic of the two-step superpotentials, therefore giving rise to
Dirac delta-function singularities in the corresponding potentials if they are
considered in the whole line $x\in(-\infty,+\infty)$.
The translational SI potentials were shown to possess a potential algebra
involving three generators of angular momentum type. The potential
algebra for the case of SI in $k$ steps ($k$ being an arbitrary positive integer)
was found \cite{Su09} to be equivalent to the generalized deformed oscillator algebra that
had a built-in $Z_{4}$ grading structure and was constructed in terms of the generators of
the deformed harmonic oscillator $(I,A,A^{\dagger},N)$ as well as the grading generator
$T$ of the cyclic group of order $k$. The obtained potentials included the cyclic SI
potentials of period $k$ as a special case.

On the other hand, the framework of $\cN$-fold SUSY \cite{AIS93,AST01b,AS03}
has gained much interest during the last few years. It is possibly one of the most
powerful frameworks among various existing methods for finding or constructing
quantum systems which admit exact solutions in a certain sense. In particular,
a few remarkable features are in order here:\\
1) $\cN$-fold SUSY includes all the ordinary SUSY methods as its
special cases and is equivalent to weak quasi-solvability \cite{AST01b} which is less
restrictive concept than quasi-exact solvability \cite{TU87,Us94}.\\
2) An arbitrary one-body quantum Hamiltonian which admits two (local) solutions
in closed form belongs to a special class of $\cN$-fold SUSY, namely, type A
2-fold SUSY \cite{AST01a,GT06} irrespective of whether or not it is Hermitian,
$\cP\cT$ symmetric \cite{BB98a}, pseudo-Hermitian \cite{Mo02a} and so on.\\
3) Many of the so far constructed (quasi-)solvable position-dependent mass
quantum Hamiltonians are also realized as special cases of type A $\cN$-fold SUSY
\cite{Ta06a}.\\
We note that almost all the models having essentially the same symmetry as
$\cN$-fold SUSY but called with other terminologies in the literature, such as
P\"{o}schl--Teller and Lam\'{e} potentials, are actually also particular cases of type
A $\cN$-fold SUSY. For a review of $\cN$-fold SUSY, see Ref.~\cite{Ta09}.
Recently in Refs.~\cite{BT09,BT10}, the necessary and sufficient conditions for
type A $\cN$-fold SUSY Hamiltonians to possess intermediate Hamiltonians were
derived for $\cN=2$ and $3$. As a by-product, some well-known translational SI
potentials were naturally obtained. It indicates that one can in principle go one
step further and consider two-step SI also in the framework of $\cN$-fold SUSY.

In this paper, we extensively investigate two-step SI in the framework of
$\cN$-fold SUSY. A key ingredient of our approach relies on the observation
that any two-step SI system possesses type A $2$-fold SUSY with an intermediate
Hamiltonian whose analytic and algebraic structures are both well understood.
The latter fact indeed enables us to make a sophisticated analysis on general aspects
of two-step SI without recourse to any specific assumption or model. Employing
the general form of type A $2$-fold SUSY, we systematically construct two-step
SI potentials without relying on any ad hoc ansatz. In addition to the well-known
ordinary SI potentials, we successfully obtain several novel two-step SI ones which
have not been reported in the existing literature. Furthermore, some of the latter
two-step SI are conditional and thus the corresponding systems are conditionally
solvable.

We organize this paper as follows. In Section~\ref{sec:tssi}, we review the concept of
two-step SI and discuss its general aspects. In particular, we show that any two-step
SI system has type A $2$-fold SUSY with an intermediate Hamiltonian and second-order
paraSUSY and that two-step SI always means solvability and actually lies between
solvability and ordinary SI. In Section~\ref{sec:t2SI}, we investigate two-step SI for
particular models which can be realized in the cases of type A $\cN$-fold SUSY for
$\cN>2$. The obtained two-step SI potentials include the well-known ones such as
(radial) harmonic oscillators, Morse, Scarf, P\"{o}schl--Teller potentials, and so on.
In addition, we also obtain some novel potentials which are expressible in terms of
elliptic, exponential (including hyperbolic and trigonometric), and rational functions.
In Section~\ref{sec:nt2SI}, we construct more general two-step SI potentials under
a less restrictive condition. We obtain more novel two-step SI potentials some of which
admit analytic expressions only implicitly. Most of the new potentials obtained in
Sections~\ref{sec:t2SI} and \ref{sec:nt2SI} do not have ordinary SI and some of their
two-step SI are conditional. Finally in Section~\ref{sec:discus},
we summarize the results and discuss perspectives and future issues.

\section{Two-step Shape Invariance and Type A $2$-fold SUSY}
\label{sec:tssi}
%\nosections%IOPART

First of all, let us introduce the concept of two-step SI.
Suppose we have a system composed of two sets of SUSY QM
\begin{align}
2H_{i}^{\pm}(\bmc)=A_{i}^{\mp}(\bmc)A_{i}^{\pm}(\bmc),\qquad
 A_{i}^{\pm}(\bmc)=\mp\frac{\rmd}{\rmd x}+W_{i}(x;\bmc),\qquad (i=0,1),
\label{eq:2susy}
\end{align}
where $\bmc$ denotes a set of all the parameters involved in the system, which
satisfies the following constraint
\begin{align}
H_{0}^{+}(\bmc)=H_{1}^{-}(\bmc)+R(\bmc),
\label{eq:inc}
\end{align}
where $R(\bmc)$ is a constant depending only on $\bmc$. According to
Ref.~\cite{BDGKPS93}, the system is said to have \emph{two-step shape
invariance} if $H_{0}^{-}$ and $H_{1}^{+}$ satisfy
\begin{align}
H_{1}^{+}(\bmc^{(0)})=H_{0}^{-}(\bmc^{(2)})+\tR_{2}(\bmc^{(0)}),
\label{eq:tssi}
\end{align}
where $\tR_{2}(\bmc^{(0)})$ is another constant and $\bmc^{(2)}=\bmc^{(2)}(\bmc^{(0)})$
is another set of parameters both of which depend only on $\bmc^{(0)}$.

To begin with, we shall show in what follows that any such a system has type A $2$-fold
SUSY with an intermediate Hamiltonian studied in Ref.~\cite{BT09}. We first note from
(\ref{eq:2susy}) and (\ref{eq:inc}) that $H_{0}^{-}(\bmc)$ and $H_{1}^{+}(\bmc)+R(\bmc)$
are intertwined by the second-order linear differential operator $A_{1}^{-}A_{0}^{-}$ as
\begin{align}
A_{1}^{-}(\bmc)A_{0}^{-}(\bmc)H_{0}^{-}(\bmc)=
 (H_{1}^{+}(\bmc)+R(\bmc))A_{1}^{-}(\bmc)A_{0}^{-}(\bmc),
\label{eq:AA}
\end{align}
and thus form a pair of $2$-fold SUSY. It is evident from (\ref{eq:inc}) that the latter
$2$-fold SUSY system possesses an intermediate Hamiltonian. In addition, we can have
an expression of the kernel of $A_{1}^{-}(\bmc)A_{0}^{-}(\bmc)$ in closed form as
\begin{align}
\ker A_{1}^{-}(\bmc)A_{0}^{-}(\bmc)=&\;\left\langle 1,\int^{x}\rmd x'\exp\biggl(\int^{x'}
 \rmd x''\,(W_{0}(x'';\bmc)-W_{1}(x'';\bmc))\biggr)\right\rangle\notag\\
&\;\times\exp\left(-\int^{x}\rmd x'\,W_{0}(x';\bmc)\right),
\end{align}
which is preserved by $H_{0}^{-}(\bmc)$ due to the intertwining relation (\ref{eq:AA}).
Hence, $H_{0}^{-}(\bmc)$ admits two linearly independent local solutions in closed form.
The latter fact is a necessary and sufficient condition for a one-dimensional Hamiltonian
to belong to type A $2$-fold SUSY proved in Ref.~\cite{GT06}. Therefore, we finally
conclude that any two-step SI system has type A $2$-fold SUSY with an intermediate
Hamiltonian.

It is now clear from the above process of the proof that any two-step SI system
(\ref{eq:2susy})--(\ref{eq:tssi}) has one-to-one correspondence with a type A $2$-fold
SUSY system with an intermediate Hamiltonian $(H^{\pm},H^{\rmi1},P_{2}^{-}=P_{21}^{-}
P_{22}^{-})$ in Ref.~\cite{BT09} by
\begin{align}
H_{0}^{-}\lra H^{-},\quad H_{0}^{+}\lra H^{\rmi1},\quad H_{1}^{+}+R\lra H^{+},
 \quad A_{0}^{\pm}\lra P_{22}^{\pm},\quad A_{1}^{\pm}\lra P_{21}^{\pm}.
\end{align}
To investigate general structure of two-step SI systems in the framework of type A
$2$-fold SUSY with an intermediate Hamiltonian, it is convenient to make a `gauge'
transformation from the physical $x$-space to a gauged $z$-space defined by
\begin{align}
\tH^{\pm}=\rme^{\cW_{2}^{-}}H^{\pm}\rme^{-\cW_{2}^{-}},\qquad
 \tP_{2}^{\pm}=\rme^{\cW_{2}^{-}}P_{2}^{\pm}\rme^{-\cW_{2}^{-}}.
\end{align}
According to Ref.~\cite{BT09}, the general form of a gauged type A $2$-fold SUSY
system with an intermediate Hamiltonian is given by
\begin{align}
\tH^{-}(\bmc)&=-A(z)\frac{\rmd^{2}}{\rmd z^{2}}-Q(z;\bmb)\frac{\rmd}{\rmd z}
 +\frac{1}{2}Q'(z;\bmb)-R,
\label{eq:tAH-}\\
\tH^{+}(\bmc)&=-A(z)\frac{\rmd^{2}}{\rmd z^{2}}-Q(z;\bmb)\frac{\rmd}{\rmd z}
 -\frac{3}{2}Q'(z;\bmb)+\frac{Q(z;\bmb)A'(z)}{A(z)}-R,
\label{eq:tAH+}\\
\tP_{2}^{-}(\bmc)&=(z')^{2}\frac{\rmd^{2}}{\rmd z^{2}},\quad\tP_{2}^{+}(\bmc)=
 \rme^{-\int\rmd z\frac{Q}{A}}\bar{P}_{2}^{+}\rme^{\int\rmd z\frac{Q}{A}}
 =(z')^{2}\left(\frac{\rmd}{\rmd z}+\frac{Q(z;\bmb)}{A(z)}\right)^{2},
\label{eq:tP2+-}
\end{align}
with the set of parameters $\bmc=\{\bmb,R\}$, where $A(z)$ is an arbitrary function
of $z$ at present and $Q(z;\bmb)$ is a polynomial of at most first-degree in $z$
\begin{align}
Q(z;\bmb)=b_{1}z+b_{0}.
\label{eq:Qz}
\end{align}
The function $z'=\rmd z/\rmd x$ which connects the variable $z$ in the gauged space
and the coordinate $x$ in the physical space is determined by
\begin{align}
z'(x)^{2}=2A(z)\bigr|_{z=z(x)}.
\label{eq:zq}
\end{align}
It can be easily checked by a direct calculation that $\tH^{\pm}$ are intertwined
by $\tP_{2}^{\pm}$ as
\begin{align}
\tP_{2}^{-}(\bmc)\tH^{-}(\bmc)=\tH^{+}(\bmc)\tP_{2}^{-}(\bmc),\qquad
 \tP_{2}^{+}(\bmc)\tH^{+}(\bmc)=\tH^{-}(\bmc)\tP_{2}^{+}(\bmc).
\label{eq:inter}
\end{align}
With a factorized form of the $2$-fold supercharges
\begin{align}
\tP_{2}^{-}(\bmc)=\tP_{21}^{-}(\bmc)\tP_{22}^{-}(\bmc),\qquad
 \tP_{2}^{+}(\bmc)=\tP_{22}^{+}(\bmc)\tP_{21}^{+}(\bmc),
\label{eq:tP2f}
\end{align}
the three component Hamiltonians $\tH^{-}$, $\tH^{\rmi1}$, and $\tH^{+}$ are written as
\begin{align}
2\tH^{-}(\bmc)&=\tP_{22}^{+}(\bmc)\tP_{22}^{-}(\bmc)+2C_{22}(\bmc),
\label{eq:tH-}\\
2\tH^{\rmi1}(\bmc)&=\tP_{22}^{-}(\bmc)\tP_{22}^{+}(\bmc)+2C_{22}(\bmc)
 =\tP_{21}^{+}(\bmc)\tP_{21}^{-}(\bmc)+2C_{21}(\bmc),
\label{eq:tHi1}\\
2\tH^{+}(\bmc)&=\tP_{21}^{-}(\bmc)\tP_{21}^{+}(\bmc)+2C_{21}(\bmc).
\label{eq:tH+}
\end{align}
The type A gauged $2$-fold supercharges (\ref{eq:tP2+-}) admit a factorization
(\ref{eq:tP2f}) with
\begin{align}
\begin{split}
\tP_{21}^{-}(\bmc)&=z'\left(\frac{\rmd}{\rmd z}-\frac{A'(z)}{2A(z)}\right),\qquad
 \tP_{22}^{-}(\bmc)=z'\frac{\rmd}{\rmd z},\\
\tP_{22}^{+}(\bmc)&=-z'\left(\frac{\rmd}{\rmd z}+\frac{2Q(z;\bmb)-A'(z)}{2A(z)}\right),
 \quad\tP_{21}^{+}(\bmc)=-z'\left(\frac{\rmd}{\rmd z}+\frac{Q(z;\bmb)}{A(z)}\right).
\end{split}
\label{eq:tP2f1}
\end{align}
It is easy to check that the pair of Hamiltonians $\tH^{\pm}$ (\ref{eq:tH-})
and (\ref{eq:tH+}) obtained from (\ref{eq:tP2f1}) actually coincides with the pair of
type A gauged Hamiltonians (\ref{eq:tAH-}) and (\ref{eq:tAH+}) with
\begin{align}
2C_{22}=b_{1}-2R,\qquad 2C_{21}=-b_{1}-2R.
\label{eq:C2i}
\end{align}
The intermediate gauged Hamiltonian $\tH^{\rmi1}$ (\ref{eq:tHi1}) with respect
to the factorization (\ref{eq:tP2f1}) is calculated as
\begin{align}
\tH^{\rmi1}(\bmc)=-A(z)\frac{\rmd^{2}}{\rmd z^{2}}-Q(z;\bmb)\frac{\rmd}{\rmd z}
 +\frac{A''(z)}{2}+\frac{(2Q(z;\bmb)-A'(z))A'(z)}{4A(z)}-\frac{b_{1}}{2}-R.
\label{eq:tAHi1}
\end{align}
As was shown in Ref.~\cite{BT09}, the type A gauged $2$-fold supercharges (\ref{eq:tP2+-})
admit, in addition to (\ref{eq:tP2f1}), a one-parameter family of factorizations, denoted by
symbols with wide hat, with
\begin{align}
\begin{split}
\htP{}_{21}^{-}&=z'\left(\frac{\rmd}{\rmd z}-\frac{A'(z)}{2A(z)}+\frac{1}{z+z_{0}}\right),
 \quad\htP{}_{22}^{-}=z'\left(\frac{\rmd}{\rmd z}-\frac{1}{z+z_{0}}\right),\\
\htP{}_{22}^{+}&=-z'\left(\frac{\rmd}{\rmd z}+\frac{2Q(z;\bmb)-A'(z)}{2A(z)}
 +\frac{1}{z+z_{0}}\right),\\
\htP{}_{21}^{+}&=-z'\left(\frac{\rmd}{\rmd z}+\frac{Q(z;\bmb)}{A(z)}-\frac{1}{z+z_{0}}\right),
\end{split}
\label{eq:tP2f2}
\end{align}
where $z_{0}\in\bbC$ is a parameter. Then, a type A $2$-fold SUSY system admits
another intermediate gauged Hamiltonian $\tH^{\rmi2}$ with respect to one of the
above one-parameter family of factorizations if and only if $b_{1}\neq0$.
In fact, if it is the case, the pair of gauged Hamiltonians $\tH^{\pm}$ (\ref{eq:tH-})
and (\ref{eq:tH+}) calculated with the second factorization (\ref{eq:tP2f2}) coincides
with the pair of type A gauged Hamiltonians (\ref{eq:tAH-}) and (\ref{eq:tAH+}) with
\begin{align}
2\widehat{C}_{22}=-b_{1}-2R,\qquad 2\widehat{C}_{21}=b_{1}-2R,
\end{align}
provided that $z_{0}=b_{0}/b_{1}$. The second intermediate gauged Hamiltonian
$\tH^{\rmi2}$ with respect to the second factorization (\ref{eq:tP2f2}) defined via
(\ref{eq:tHi1}) with the corresponding hatted quantities is calculated as
\begin{align}
\tH^{\rmi2}=\tH^{\rmi1}-\frac{b_{1}A'(z)}{Q(z;\bmb)}+\frac{2(b_{1})^{2}A(z)}{Q(z;\bmb)^{2}}.
\label{eq:tAHi2}
\end{align}
Another remarkable aspect of type A $2$-fold SUSY systems with an intermediate
Hamiltonian is that they possess second-order paraSUSY as well~\cite{BT09}. Introducing
a triple of operators $(\bH_{\rmP}, \bQ_{\rmP}^{\pm})$ as
\begin{align}
\begin{split}
&\bH_{\rmP}=H^{-}(\psi_{\rmP}^{-})^{2}(\psi_{\rmP}^{+})^{2}+H^{\rmi1}(\psi_{\rmP}^{+}
 \psi_{\rmP}^{-}-(\psi_{\rmP}^{+})^{2}(\psi_{\rmP}^{-})^{2})+H^{+}(\psi_{\rmP}^{+})^{2}
 (\psi_{\rmP}^{-})^{2},\\
&\bQ_{\rmP}^{-}=P_{22}^{+}(\psi_{\rmP}^{-})^{2}\psi_{\rmP}^{+}+P_{21}^{+}\psi_{\rmP}^{+}
 (\psi_{\rmP}^{-})^{2},\qquad \bQ_{\rmP}^{+}=P_{22}^{-}\psi_{\rmP}^{-}(\psi_{\rmP}^{+})^{2}
 +P_{21}^{-}(\psi_{\rmP}^{+})^{2}\psi_{\rmP}^{-},
\end{split}
\end{align}
where $H^{\pm}$ are a pair of type A $2$-fold SUSY Hamiltonians, $H^{\rmi1}$ is
(one of) its intermediate Hamiltonians, and $\psi_{\rmP}^{\pm}$ are second-order
parafermions satisfying
\begin{align}
(\psi_{\rmP}^{\pm})^{2}\neq 0,\qquad (\psi_{\rmP}^{\pm})^{3}=0,\qquad \bigl\{
 \psi_{\rmP}^{-},\psi_{\rmP}^{+}\bigr\}+\bigl\{ (\psi_{\rmP}^{-})^{2}, (\psi_{\rmP}^{+})^{2}
 \bigr\}=2 I,
\end{align}
we see that the triple $(\bH_{\rmP}, \bQ_{\rmP}^{\pm})$ satisfies the second-order
paraSUSY relations in \cite {RS88}
\begin{align}
\begin{split}
(\bQ_{\rmP}^{\pm})^{2}\neq 0,\qquad (\bQ_{\rmP}^{\pm})^{3}=0,\qquad \bigl[
 \bQ_{\rmP}^{\pm}, \bH_{\rmP}\bigr] =0,\\
(\bQ_{\rmP}^{\pm})^{2}\bQ_{\rmP}^{\mp}+\bQ_{\rmP}^{\pm}\bQ_{\rmP}^{\mp}
 \bQ_{\rmP}^{\pm}+\bQ_{\rmP}^{\mp}(\bQ_{\rmP}^{\pm})^{2}=4\bQ_{\rmP}^{\pm}\bH_{\rmP},
\end{split}
\end{align}
as well as the generalized $2$-fold superalgebra
\begin{align}
(\bQ_{\rmP}^{-})^{2}(\bQ_{\rmP}^{+})^{2}+\bQ_{\rmP}^{\pm}(\bQ_{\rmP}^{\mp})^{2}
 \bQ_{\rmP}^{\pm}+(\bQ_{\rmP}^{+})^{2}(\bQ_{\rmP}^{-})^{2}=4(\bH_{\rmP})^{2}-(b_{1})^{2},
\end{align}
where $b_{1}$ is the same parameter as the one appeared in (\ref{eq:Qz}).

To see what are significant consequences of two-step SI, let us first show that if a pair
of type A $2$-fold SUSY gauged Hamiltonians $\tH^{\pm}$ satisfies the two-step SI
condition (\ref{eq:tssi}), that is,
\begin{align}
\tH^{+}(\bmc^{(2k)})=\tH^{-}(\bmc^{(2k+2)})+\tR_{2}(\bmc^{(2k)}),
\label{eq:tssi2}
\end{align}
where we have generalized $\bmc^{(0)}$ and $\bmc^{(2)}$ in (\ref{eq:tssi}) to $\bmc^{(2k)}$
and $\bmc^{(2k+2)}$ ($k=0,1,2,\ldots$), and if, in addition, $\tH^{-}(\bmc^{(2k+2)})$
preserves a linear space $\tcV^{-}(\bmc^{(2k+2)})$, then $\tH^{-}(\bmc^{(2k)})$ preserves
the space $\tP_{2}^{+}(\bmc^{(2k)})\tcV^{-}(\bmc^{(2k+2)})$. In a mathematical language,
we have
\begin{align}
&\tH^{-}(\bmc^{(2k+2)})\tcV^{-}(\bmc^{(2k+2)})\subset\tcV^{-}(\bmc^{(2k+2)})\notag\\
&\ \Longrightarrow\ \tH^{-}(\bmc^{(2k)})\tP_{2}^{+}(\bmc^{(2k)})\tcV^{-}(\bmc^{(2k+2)})
 \subset\tP_{2}^{+}(\bmc^{(2k)})\tcV^{-}(\bmc^{(2k+2)}).
\label{eq:1}
\end{align}
Indeed, it follows from (\ref{eq:tP2f})--(\ref{eq:tH+}) that
\begin{align*}
\lefteqn{
2\tH^{-}(\bmc^{(2k)})\tP_{2}^{+}(\bmc^{(2k)})\tcV^{-}(\bmc^{(2k+2)})
}\\
&=\left[\tP_{22}^{+}(\bmc^{(2k)})\tP_{22}^{-}(\bmc^{(2k)})+2C_{22}(\bmc^{(2k)})\right]
 \tP_{22}^{+}(\bmc^{(2k)})\tP_{21}^{+}(\bmc^{(2k)})\tcV^{-}(\bmc^{(2k+2)})\\
&=\tP_{22}^{+}(\bmc^{(2k)})\left[\tP_{22}^{-}(\bmc^{(2k)})\tP_{22}^{+}(\bmc^{(2k)})
 +2C_{22}(\bmc^{(2k)})\right]\tP_{21}^{+}(\bmc^{(2k)})\tcV^{-}(\bmc^{(2k+2)})\\
&=\tP_{22}^{+}(\bmc^{(2k)})\left[\tP_{21}^{+}(\bmc^{(2k)})\tP_{21}^{-}(\bmc^{(2k)})
 +2C_{21}(\bmc^{(2k)})\right]\tP_{21}^{+}(\bmc^{(2k)})\tcV^{-}(\bmc^{(2k+2)})\\
&=\tP_{22}^{+}(\bmc^{(2k)})\tP_{21}^{+}(\bmc^{(2k)})\left[\tP_{21}^{-}(\bmc^{(2k)})
 \tP_{21}^{+}(\bmc^{(2k)})+2C_{21}(\bmc^{(2k)})\right]\tcV^{-}(\bmc^{(2k+2)})\\
&=2\tP_{2}^{+}(\bmc^{(2k)})\tH^{+}(\bmc^{(2k)})\tcV^{-}(\bmc^{(2k+2)}).
\end{align*}
Hence, under the condition (\ref{eq:tssi2}) we obtain
\begin{align}
\tH^{-}(\bmc^{(2k)})\tP_{2}^{+}(\bmc^{(2k)})\tcV^{-}(\bmc^{(2k+2)})=\tP_{2}^{+}(\bmc^{(2k)})
 \left[\tH^{-}(\bmc^{(2k+2)})+\tR_{2}(\bmc^{(2k)})\right]\tcV^{-}(\bmc^{(2k+2)}),
\end{align}
and thus conclude (\ref{eq:1}). On the other hand, it is evident from (\ref{eq:inter})
that $\tH^{-}$ preserves the kernel of $\tP_{2}^{-}$ which is at most two dimensional:
\begin{align}
\tH^{-}(\bmc)\tcV_{2}^{-}(\bmc)\subset\tcV_{2}^{-}(\bmc),\qquad
 \tcV_{2}^{-}(\bmc)=\ker\tP_{2}^{-}(\bmc)=\braket{1,z},
\label{eq:2}
\end{align}
and thus it is \emph{quasi-solvable}~\cite{Ta09}.
Combining (\ref{eq:1}) and (\ref{eq:2}), we see that the Hamiltonian
$\tH^{-}(\bmc^{(0)})$ preserves an infinite flag of finite-dimensional linear spaces
\begin{align}
\tcV_{2}^{-}(\bmc^{(0)})\subset\tcV_{4}^{-}(\bmc^{(0)})\subset\dots\subset
 \tcV_{2n}^{-}(\bmc^{(0)})\subset\cdots,
\end{align}
where $\tcV_{2n}^{-}(\bmc^{(0)})$ is defined by
\begin{align}
\tcV_{2n}^{-}(\bmc^{(0)})=&\;\tcV_{2}^{-}(\bmc^{(0)})+\tP_{2}^{+}(\bmc^{(0)})\tcV_{2}^{-}
 (\bmc^{(2)})+\cdots\notag\\
&\;\cdots+\tP_{2}^{+}(\bmc^{(0)})\tP_{2}^{+}(\bmc^{(2)})\cdots\tP_{2}^{+}(\bmc^{(2n-4)})
 \tcV_{2}^{-}(\bmc^{(2n-2)}).
\end{align}
Hence, any two-step SI Hamiltonian is \emph{solvable}~\cite{Ta09} unless there is
a natural number $n$ such that $\tcV_{2n}^{-}(\bmc^{(0)})\subset\tcV_{2}^{-}(\bmc^{(0)})$.
We also note that if an intermediate Hamiltonian $\tH^{\rmi1}$ satisfies
\begin{align}
\tH^{\rmi1}(\bmc^{(2k+1)})=\tH^{-}(\bmc^{(2k+2)})+\tR_{1}(\bmc^{(2k+1)}),
\end{align}
where $\tR_{1}(\bmc^{(2k+1)})$ is a constant depending only on $\bmc^{(2k+1)}$, then
the three Hamiltonians $\tH^{-}$, $\tH^|{\rmi1}$, and $\tH^{+}$ constitute
a sequence of ordinary SI with
\begin{align}
\tH^{+}(\bmc^{(2k)})=\tH^{\rmi1}(\bmc^{(2k+1)})+\tR_{1}(\bmc^{(2k)}),
\end{align}
where $\tR_{2}(\bmc^{(2k)})=\tR_{1}(\bmc^{(2k+1)})+\tR_{1}(\bmc^{(2k)})$. Conversely,
it is evident from the definition that any ordinary SI Hamiltonian is two-step SI.
It means that ordinary SI can be regarded as a sufficient condition for two-step SI.
Summarizing the above arguments, we have the following relation:
\begin{align*}
\text{(Solvability)}\supset\text{(Two-step SI)}\supset\text{(Ordinary SI)}.
\end{align*}
We shall say that two-step SI is \emph{irreducible} if it is \emph{not} ordinary SI
simultaneously, and is \emph{reducible} otherwise.

The pair of type A 2-fold SUSY potentials $V^{\pm}$ with an intermediate
Hamiltonian is given by~\cite{BT09},
\begin{align}
V^{\pm}(x,\bmc)=&-\frac{1}{4A(z)}\biggl[A(z)A''(z)-\frac{3}{4}A'(z)^{2}
 -(b_{1}z+b_{0})^{2}\notag\\
&\mp 2\bigl((b_{1}z+b_{0})A'(z)-2b_{1}A(z)\bigr)\biggr]\biggr|_{z=z(x)}-R,
\label{eq:V+-}
\end{align}
where the set of parameters is $\bmc=\{b_{1},b_{0},R\}$. The potential terms in
the first and second intermediate Hamiltonians $H^{\rmi1}$ and $H^{\rmi2}$,
the latter of which exists only when $b_{1}\neq0$, are calculated as
\begin{align}
V^{\rmi1}(x,\bmc)&=\frac{A''(z)}{4}-\frac{A'(z)^{2}-4(b_{1}z+b_{0})^{2}}{16A(z)}
 \biggr|_{z=z(x)}-R,
\label{eq:Vi1}\\
V^{\rmi2}(x;\bmc)&=V^{\rmi1}(x;\bmc)-\frac{b_{1}A'(z)}{b_{1}z+b_{0}}
 +\frac{2(b_{1})^{2}A(z)}{(b_{1}z+b_{0})^{2}}\biggr|_{z=z(x)}-R.
\label{eq:Vi2}
\end{align}
In what follows, we put $R=0$ to remove an irrelevant parameter without any
loss of generality. From the form of a pair of potentials (\ref{eq:V+-}), we immediately
see that there is always two-step SI of the following type
\begin{align}
V^{+}(x;b_{1},b_{0})=V^{-}(x;-b_{1},-b_{0}),
\label{eq:refts}
\end{align}
that is, the relation (\ref{eq:tssi}) holds with $b_{1}^{(2)}=-b_{1}^{(0)}$, $b_{0}^{(2)}=
-b_{0}^{(0)}$, and $\tR_{2}(b_{1}^{(0)},b_{0}^{(0)})=0$, irrespective of the form of $A(z)$.
The latter two-step SI usually changes the normalizability of the corresponding
solvable sectors and thus can be employed to convert a solvable model without
normalizable eigenfunctions into an exactly solvable one. A similar tactics was
already demonstrated, although with a different type from (\ref{eq:refts}),
in Ref.~\cite{GMS01}. We shall say that the two-step SI under consideration is
\emph{reflective} if it is characterized by (\ref{eq:refts}). In general, reflective
two-step SI is irreducible. Our central concern is now what kinds of $A(z)$ can
admit non-reflective two-step SI.

\section{Polynomial $A(z)$}
\label{sec:t2SI}

To begin with, we recall the fact that in type A $\cN$-fold SUSY for $\cN>2$
the function $A(z)$ must be a polynomial of at most fourth degree:
\begin{align}
A(z)=a_{4}z^{4}+a_{3}z^{3}+a_{2}z^{2}+a_{1}z+a_{0}.
\end{align}
Hence, it is natural to examine first the case of this polynomial $A(z)$ of at most
fourth degree. In the followings, we present the forms of potentials resulting from
a polynomial $A(z)$ of from zeroth to fourth degree.\\

\noindent
\textbf{Case 1-1:} $a_{4}=a_{3}=a_{2}=a_{1}=0$ \& $a_{0}\neq0$\\
Let us first consider the case of a zeroth degree $A(z)$. In this case, the pair of
potentials (\ref{eq:V+-}) are given by
\begin{align}
V^{\pm}(x;b_{1},b_{0})=\frac{(b_{1})^{2}z^{2}+2b_{1}b_{0}z+(b_{0})^{2}\mp
 4a_{0}b_{1}}{4a_{0}}\biggr|_{z=z(x)}.
\label{eq:C11V}
\end{align}
In addition to (\ref{eq:refts}), they have non-reflective two-step SI as
\begin{align}
V^{+}(x;b_{1},b_{0})=V^{-}(x;b_{1},b_{0})-2b_{1},
\end{align}
that is, the relation (\ref{eq:tssi}) holds with $b_{1}^{(2)}=b_{1}^{(0)}$, $b_{0}^{(2)}=
b_{0}^{(0)}$, and $\tR_{2}(b_{1}^{(0)},b_{0}^{(0)})=-2b_{1}^{(0)}$. The first intermediate
potential (\ref{eq:Vi1}) is calculated as
\begin{align}
V^{\rmi1}(x;b_{1},b_{0})=\frac{(b_{1})^{2}z^{2}+2b_{1}b_{0}z+(b_{0})^{2}}{4a_{0}}\biggr|_{z=z(x)},
\label{eq:C11i1}
\end{align}
and thus it together with $V^{\pm}$ constitutes a sequence of ordinary SI as
\begin{align}
\begin{split}
V^{\rmi1}(x;b_{1},b_{0})=V^{-}(x;b_{1},b_{0})-b_{1},\\
V^{+}(x;b_{1},b_{0})=V^{\rmi1}(x;b_{1},b_{0})-b_{1}.
\end{split}
\end{align}
Therefore, the system has reducible two-step SI. On the other hand, the second
intermediate potential (\ref{eq:Vi2}) which exists only when $b_{1}\neq0$ reads as
\begin{align}
V^{\rmi2}(x;b_{1},b_{0})=V^{\rmi1}(x;b_{1},b_{0})+\frac{2a_{0}(b_{1})^{2}}{(b_{1}z+b_{0})^{2}}
 \biggr|_{z=z(x)},
\label{eq:C11i2}
\end{align}
and has no SI with the other potentials. The equation (\ref{eq:zq}) in the present case
is integrated as
\begin{align}
z=\sqrt{2a_{0}}\,x.
\label{eq:C11z}
\end{align}
Substituting (\ref{eq:C11z}) into (\ref{eq:C11V}), (\ref{eq:C11i1}), and (\ref{eq:C11i2}),
we finally obtain the potentials in the $x$-space. The system is well-known harmonic
oscillators.\\

\noindent
\textbf{Case 1-2:} $a_{4}=a_{3}=a_{2}=0$ \& $a_{1}\neq0$

Next, we shall consider a first degree $A(z)$ with $a_{4}=a_{3}=a_{2}=0$ and $a_{1}\neq0$.
In this case, the pair of potentials (\ref{eq:V+-}) are given by
\begin{align}
V^{\pm}(x;b_{1},b_{0})=&\;\frac{(b_{1})^{2}}{4a_{1}}z+\frac{1}{4(a_{1}z+a_{0})}\left(
 \frac{b_{1}a_{0}}{a_{1}}-\frac{2b_{0}\pm 3a_{1}}{2}\right)\left(\frac{b_{1}a_{0}}{a_{1}}
 -\frac{2b_{0}\pm a_{1}}{2}\right)\notag\\
&\;-\frac{b_{1}}{4a_{1}}\left(\frac{b_{1}a_{0}}{a_{1}}-2b_{0}\pm 2a_{1}\right)
 \biggr|_{z=z(x)}.
\label{eq:C12V}
\end{align}
In addition to (\ref{eq:refts}), they have non-reflective two-step SI as
\begin{align}
V^{+}(x;b_{1},b_{0})=V^{-}(x;b_{1},b_{0}+2a_{1})-2b_{1},
\end{align}
that is, the relation (\ref{eq:tssi}) holds with $b_{1}^{(2)}=b_{1}^{(0)}$, $b_{0}^{(2)}=
b_{0}^{(0)}+2a_{1}$, and $\tR_{2}(b_{1}^{(0},b_{0}^{(0)})=-2b_{1}^{(0)}$. The first intermediate
potential (\ref{eq:Vi1}) is calculated as
\begin{align}
V^{\rmi1}(x;b_{1},b_{0})=&\;\frac{(b_{1})^{2}}{4a_{1}}z+\frac{1}{4(a_{1}z+a_{0})}\left(
 \frac{b_{1}a_{0}}{a_{1}}-\frac{2b_{0}+a_{1}}{2}\right)\left(\frac{b_{1}a_{0}}{a_{1}}
 -\frac{2b_{0}-a_{1}}{2}\right)\notag\\
&\;-\frac{b_{1}}{4a_{1}}\left(\frac{b_{1}a_{0}}{a_{1}}-2b_{0}\right)\biggr|_{z=z(x)},
\label{eq:C12i1}
\end{align}
and thus it together with $V^{\pm}$ constitutes a sequence of ordinary SI as
\begin{align}
\begin{split}
V^{\rmi1}(x;b_{1},b_{0})=V^{-}(x;b_{1},b_{0}+a_{1})-b_{1},\\
V^{+}(x;b_{1},b_{0})=V^{\rmi1}(x;b_{1},b_{0}+a_{1})-b_{1}.
\end{split}
\end{align}
Therefore, the system has reducible two-step SI. On the other hand, the second
intermediate potential (\ref{eq:Vi2}) which exists only when $b_{1}\neq0$ reads as
\begin{align}
V^{\rmi2}(x;b_{1},b_{0})=V^{\rmi1}(x;b_{1},b_{0})+\frac{a_{1}b_{1}}{b_{1}z+b_{0}}
 +\frac{2(a_{0}b_{1}-a_{1}b_{0})b_{1}}{(b_{1}z+b_{0})^{2}}\biggr|_{z=z(x)},
\label{eq:C12i2}
\end{align}
and has no SI with the other potentials. The equation (\ref{eq:zq}) in the present case
is integrated as
\begin{align}
z=\frac{a_{1}}{2}x^{2}-\frac{a_{0}}{a_{1}}.
\label{eq:C12z}
\end{align}
Substituting (\ref{eq:C12z}) into (\ref{eq:C12V}), (\ref{eq:C12i1}), and (\ref{eq:C12i2}),
we finally obtain the potentials in the $x$-space. The system consists of well-known
radial harmonic oscillators.\\

\noindent
\textbf{Case 1-3:} $a_{4}=a_{3}=0$ \& $a_{2}\neq0$

Next, we shall consider the case of a second degree $A(z)$.
In this case, the pair of potentials (\ref{eq:V+-}) are given by
\begin{align}
V^{\pm}(x;b_{1},b_{0})=&\;\frac{(b_{1})^{2}+(a_{2})^{2}}{4a_{2}}+\frac{(2a_{2}b_{0}
 -a_{1}b_{1})(b_{1}\pm 2a_{2})}{4a_{2}(a_{2}z^{2}+a_{1}z+a_{0})}z\notag\\
&\;+\frac{a_{2}(2b_{0}\pm a_{1})(2b_{0}\pm 3a_{1})-4a_{0}(b_{1}\pm a_{2})(b_{1}
 \pm 3a_{2})}{16a_{2}(a_{2}z^{2}+a_{1}z+a_{0})}\biggr|_{z=z(x)}.
\label{eq:C13V}
\end{align}
In addition to (\ref{eq:refts}), they have non-reflective two-step SI as
\begin{align}
V^{+}(x;b_{1},b_{0})=V^{-}(x;b_{1}+4a_{2},b_{0}+2a_{1})-2(b_{1}+2a_{2}),
\end{align}
that is, the relation (\ref{eq:tssi}) holds with $b_{1}^{(2)}=b_{1}^{(0)}+4a_{2}$, $b_{0}^{(2)}=
b_{0}^{(0)}+2a_{1}$, and $\tR_{2}(b_{1}^{(0},b_{0}^{(0)})=-2(b_{1}^{(0)}+2a_{2})$. The first
intermediate potential (\ref{eq:Vi1}) is calculated as
\begin{align}
V^{\rmi1}(x;b_{1},b_{0})=&\;\frac{(b_{1})^{2}+(a_{2})^{2}}{4a_{2}}+\frac{(2a_{2}b_{0}
 -a_{1}b_{1})b_{1}}{4a_{2}(a_{2}z^{2}+a_{1}z+a_{0})}z\notag\\
&\;+\frac{a_{2}(2b_{0}+a_{1})(2b_{0}-a_{1})-4a_{0}(b_{1}+a_{2})(b_{1}-a_{2})}{16a_{2}
 (a_{2}z^{2}+a_{1}z+a_{0})}\biggr|_{z=z(x)},
\label{eq:C13i1}
\end{align}
and thus it together with $V^{\pm}$ constitutes a sequence of ordinary SI as
\begin{align}
\begin{split}
V^{\rmi1}(x;b_{1},b_{0})=V^{-}(x;b_{1}+2a_{2},b_{0}+a_{1})-b_{1}-a_{2},\\
V^{+}(x;b_{1},b_{0})=V^{\rmi1}(x;b_{1}+2a_{2},b_{0}+a_{1})-b_{1}-a_{2}.
\end{split}
\end{align}
Therefore, the system has reducible two-step SI. On the other hand, the second
intermediate potential (\ref{eq:Vi2}) which exists only when $b_{1}\neq0$ reads as
\begin{align}
V^{\rmi2}(x;b_{1},b_{0})=V^{\rmi1}(x;b_{1},b_{0})+\frac{a_{1}b_{1}-2a_{2}b_{0}}{b_{1}z+b_{0}}
 +2\frac{a_{0}(b_{1})^{2}-a_{1}b_{1}b_{0}+a_{2}(b_{0})^{2}}{(b_{1}z+b_{0})^{2}}\biggr|_{z=z(x)},
\label{eq:C13i2}
\end{align}
and has no SI with the other potentials. The equation (\ref{eq:zq}) in the present case
is integrated as
\begin{align}
z+\frac{a_{1}}{2a_{2}}=\frac{1}{2}\,\rme^{\sqrt{2a_{2}}\,x}-\left(\frac{a_{0}}{2a_{2}}
 -\frac{(a_{1})^{2}}{8(a_{2})^{2}}\right)\rme^{-\sqrt{2a_{2}}\,x}.
\label{eq:C13z}
\end{align}
Substituting (\ref{eq:C13z}) into (\ref{eq:C13V}), (\ref{eq:C13i1}), and (\ref{eq:C13i2}),
we finally obtain the potentials in the $x$-space. The system consists of well-known
Morse, Scarf, or P\"{o}schl--Teller potentials, depending on the relative relation among
given values of the parameters $a_{i}$ ($i=0,1,2$).\\

\noindent
\textbf{Case 1-4:} $a_{4}=0$ \& $a_{3}\neq0$

Next, we shall consider the case of a third degree $A(z)$.
In this case, the pair of potentials (\ref{eq:V+-}) are given by
\begin{align}
V^{\pm}(x;b_{1},b_{0})=&\;\frac{3a_{3}}{16}z+\frac{a_{2}\pm8b_{1}}{16}
 -\frac{3a_{3}(3a_{1}\mp8b_{0})-a_{2}(3a_{2}\mp8b_{1})-4(b_{1})^{2}}{16A(z)}z^{2}\notag\\
&\;-\frac{27a_{3}a_{0}-a_{1}(3a_{2}\mp16b_{1})-8b_{0}(b_{1}\pm2a_{2})}{16A(z)}z\notag\\
&\;-\frac{a_{0}(9a_{2}\pm 24b_{1})-a_{1}(3a_{1}\pm8b_{0})-4(b_{0})^{2}}{
 16A(z)}\biggr|_{z=z(x)}.
\label{eq:C14V}
\end{align}
The first intermediate potential (\ref{eq:Vi1}) is calculated as
\begin{align}
V^{\rmi1}(x;b_{1},b_{0})=&\;\frac{15a_{3}}{16}z+\frac{5a_{2}}{16}+\frac{4(b_{1})^{2}
 +3a_{3}a_{1}-(a_{2})^{2}}{16A(z)}z^{2}\notag\\
&\;+\frac{8b_{1}b_{0}+9a_{3}a_{0}-a_{2}a_{1}}{16A(z)}z
 +\frac{4(b_{0})^{2}+3a_{2}a_{0}-(a_{1})^{2}}{16A(z)}\biggr|_{z=z(x)}.
\label{eq:C14i1}
\end{align}
On the other hand, the second intermediate potential (\ref{eq:Vi2}) which exists only
when $b_{1}\neq0$ reads as
\begin{align}
V^{\rmi2}(x;b_{1},b_{0})=&\;V^{\rmi1}(x;b_{1},b_{0})-a_{3}z-\frac{a_{3}b_{0}}{b_{1}}
 +\frac{a_{1}(b_{1})^{2}-2a_{2}b_{1}b_{0}+3a_{3}(b_{0})^{2}}{b_{1}(b_{1}z+b_{0})}\notag\\
&\;+2\frac{a_{0}(b_{1})^{3}-a_{1}(b_{1})^{2}b_{0}+a_{2}b_{1}(b_{0})^{2}-a_{3}(b_{0})^{3}}{
 b_{1}(b_{1}z+b_{0})^{2}}\biggr|_{z=z(x)}.
\label{eq:C14i2}
\end{align}
Although $V^{+}$ and $V^{-}$ have the same functional dependence on the variable
$z$, it does not necessarily mean that there is a transformation of the parameters $b_{1}$
and $b_{0}$ which lead to non-reflective two-step SI. In fact, they satisfy the condition for
two-step SI (\ref{eq:tssi}) if and only if
\begin{align}
\left(\begin{array}{ll} -(b_{1}^{-}+2a_{2})\ & 6a_{3}\\ b_{0}^{-}+4a_{1} & b_{1}^{-}-4a_{2}\\
 6a_{0} & b_{0}^{-}-2a_{1}\end{array}\right)\left(\begin{array}{ll} b_{1}^{+}\\ b_{0}^{+}
 \end{array}\right)=0,
\label{eq:C140}
\end{align}
where $b_{i}^{\pm}=b_{i}^{(2)}\pm b_{i}^{(0)}$. The trivial solution $b_{1}^{+}=b_{0}^{+}=0$
just produces reflective two-step SI (\ref{eq:refts}). To have a non-trivial solution,
the parameters must satisfy
\begin{align}
&6a_{3}(b_{0}^{-}+4a_{1})=-(b_{1}^{-}+2a_{2})(b_{1}^{-}-4a_{2}),
\label{eq:C141}\\
&(b_{1}^{-}+2a_{2})(b_{0}^{-}-2a_{1})=-36a_{3}a_{0}.
\label{eq:C142}
\end{align}
We first consider the particular case when $b_{1}^{-}=-2a_{2}$. In the latter case, the set
of equations (\ref{eq:C141}) and (\ref{eq:C142}) reduces by the assumption $a_{3}\neq0$ to
\begin{align}
b_{0}^{+}=b_{0}^{-}+4a_{1}=a_{0}=0.
\end{align}
Hence, we have
\begin{align}
A(z)=a_{3}z^{3}+a_{2}z^{2}+a_{1}z,\quad b_{1}^{(2)}=b_{1}^{(0)}-2a_{2},\quad
 b_{0}^{(2)}=-b_{0}^{(0)}=-2a_{1}.
\label{eq:C144}
\end{align}
In this case, the pair of potentials (\ref{eq:C14V}) reads as
\begin{align}
V^{\pm}(x;b_{1},b_{0})=&\;\frac{3a_{3}}{16}z+\frac{a_{2}\pm8b_{1}}{16}
 -\frac{3a_{3}(3a_{1}\mp8b_{0})-a_{2}(3a_{2}\mp8b_{1})-4(b_{1})^{2}}{16(a_{3}z^{2}+a_{2}z
 +a_{1})}z\notag\\
&\;+\frac{a_{1}(3a_{2}\mp16b_{1})+8b_{0}(b_{1}\pm2a_{2})}{16(a_{3}z^{2}
 +a_{2}z+a_{1})}+\frac{a_{1}(3a_{1}\pm8b_{0})+4(b_{0})^{2}}{16(a_{3}z^{2}+a_{2}z+a_{1})z}
 \biggr|_{z=z(x)}.
\label{eq:C14V'}
\end{align}
We can easily check that the latter pair satisfies
\begin{align}
V^{+}(x;b_{1},b_{0})A(z)=&\;V^{-}(x;b_{1}-2a_{2},-b_{0})A(z)\notag\\
&\;+(b_{1}-a_{2})\left[a_{3}z^{3}+a_{2}z^{2}+(b_{0}-a_{1})z\right],
\end{align}
and thus has two-step SI if and only if $b_{0}=2a_{1}$, as has been indicated by
the second equality in (\ref{eq:C144}). It is quite intriguing that a system has SI
only when a parameter fulfills a particular condition. This situation is reminiscent
of \emph{conditional exact solvability}~\cite{SoDu93}. Indeed, as was proved in
the previous section, two-step SI always means solvability, and thus the present
system is solvable only when $b_{0}=2a_{1}$. That is, it is conditionally solvable.
Hence, we shall call such SI \emph{conditional}.

The first and second intermediate potentials (\ref{eq:C14i1}) and (\ref{eq:C14i2})
in this case of conditional two-step SI read respectively as
\begin{align}
V^{\rmi1}(x;b_{1},b_{0})=&\;\frac{15a_{3}}{16}z+\frac{5a_{2}}{16}+\frac{4(b_{1})^{2}
 +3a_{3}a_{1}-(a_{2})^{2}}{16(a_{3}z^{2}+a_{2}z+a_{1})}z\notag\\
&\;+\frac{8b_{1}b_{0}-a_{2}a_{1}}{16(a_{3}z^{2}+a_{2}z+a_{1})}
 +\frac{4(b_{0})^{2}-(a_{1})^{2}}{16(a_{3}z^{2}+a_{2}z+a_{1})z}\biggr|_{z=z(x)}.
\label{eq:C14i1'}
\end{align}
and
\begin{align}
V^{\rmi2}(x;b_{1},b_{0})=&\;V^{\rmi1}(x;b_{1},b_{0})-a_{3}z-\frac{a_{3}b_{0}}{b_{1}}
 +\frac{a_{1}(b_{1})^{2}-2a_{2}b_{1}b_{0}+3a_{3}(b_{0})^{2}}{b_{1}(b_{1}z+b_{0})}\notag\\
&\;-2b_{0}\frac{a_{1}(b_{1})^{2}-a_{2}b_{1}b_{0}+a_{3}(b_{0})^{2}}{
 b_{1}(b_{1}z+b_{0})^{2}}\biggr|_{z=z(x)}.
\label{eq:C14i2'}
\end{align}
Both of them do not have ordinary SI with $V^{\pm}$ given in (\ref{eq:C14V'})
irrespective of whether $b_{0}=2a_{1}$ or not.
Hence, the conditional two-step SI of the system (\ref{eq:C14V'}) is irreducible.

The general solution to equation (\ref{eq:zq}) with $A(z)$ given in (\ref{eq:C144})
is expressed in terms of elliptic functions when $a_{1}\neq0$. The potentials are
thus of deformed Lam\'{e} or Heun type. We omit its involved general form here. When
$a_{1}=0$, the system automatically has two-step SI since $b_{0}^{(2)}=b_{0}^{(0)}=0$
from the third equality in (\ref{eq:C144}). In addition, the form of $A(z)$ reduces
to the one given in (\ref{eq:C14A''}) with $d_{1}=-a_{2}$, and the solution to
(\ref{eq:zq}) is given by (\ref{eq:C14z}). Substituting (\ref{eq:C14z}) into
(\ref{eq:C14V'}), (\ref{eq:C14i1'}), and (\ref{eq:C14i2'}), we finally obtain
the potentials in the $x$-space.\\

When $b_{1}^{-}\neq-2a_{2}$, we see from (\ref{eq:C141}) and (\ref{eq:C142}) that
$b_{1}^{-}$ and $b_{0}^{-}$ are functions of $a_{i}$ ($i=0,\dots,3$) and thus are
denoted by
\begin{align}
b_{1}^{-}=b_{1}^{(2)}-b_{1}^{(0)}=2d_{1}(a),\qquad b_{0}=b_{0}^{(2)}-b_{0}^{(0)}=2d_{0}(a).
\label{eq:C14b1}
\end{align}
Substituting (\ref{eq:C14b1}) into a general solution $6a_{3}b_{0}^{+}=(b_{1}^{-}+2a_{2})
b_{1}^{+}$ to the homogeneous equation (\ref{eq:C140}), we obtain
\begin{align}
b_{0}^{(0)}=\frac{a_{2}+d_{1}}{3a_{3}}(b_{1}^{(0)}+d_{1})-d_{0},
\label{eq:C14b2}
\end{align}
where and hereafter we omit the argument of $d_{i}$ for the brevity. The equality
(\ref{eq:C14b2}) shows that $b_{1}^{(0)}$ and $b_{0}^{(0)}$ are dependent parameters.
Then, $b_{1}^{(2)}$ and $b_{0}^{(2)}$ must fulfill the same relation as theirs for the
consistency. The latter requirement results in
\begin{align}
d_{0}=\frac{a_{2}+d_{1}}{3a_{3}}d_{1},\qquad b_{0}^{(2)}=\frac{a_{2}+d_{1}}{3a_{3}}b_{1}^{(2)},
 \qquad b_{0}^{(0)}=\frac{a_{2}+d_{1}}{3a_{3}}b_{1}^{(0)}.
\label{eq:C14b3}
\end{align}
Substituting (\ref{eq:C14b3}) back into (\ref{eq:C141}) and (\ref{eq:C142}), we have
\begin{align}
a_{1}=\frac{(a_{2})^{2}-(d_{1})^{2}}{3a_{3}},\qquad a_{0}=\frac{(a_{2}+d_{1})^{2}
 (a_{2}-2d_{1})}{27(a_{3})^{2}}.
\label{eq:C14b4}
\end{align}
Under the latter parameter relations, the function $A(z)$ is factorized as
\begin{align}
A(z)=a_{3}\left(z+\frac{a_{2}+d_{1}}{3a_{3}}\right)^{2}\left(z+\frac{a_{2}-2d_{1}}{3a_{3}}
 \right),
\label{eq:C14A''}
\end{align}
and the pair of potentials (\ref{eq:C14V}) reads as
\begin{align}
V^{\pm}(x;b_{1}):=&\;V^{\pm}\Bigl(x;b_{1},\frac{a_{2}+d_{1}}{3a_{3}}b_{1}\Bigr)
 \biggr|_{a_{1}=\frac{(a_{2})^{2}-(d_{1})^{2}}{3a_{3}},\ a_{0}=\frac{(a_{2}+d_{1})^{2}
 (a_{2}-2d_{1})}{27(a_{3})^{2}}}\notag\\
=&\;\frac{3a_{3}}{16}z+\frac{a_{2}\pm 8b_{1}}{16}+\frac{3(d_{1}\pm 2b_{1})(3d_{1}\pm
 2b_{1})}{16(3a_{3}z+a_{2}-2d_{1})}\biggr|_{z=z(x)}.
\label{eq:C14V''}
\end{align}
We now easily check that the latter pair of potentials satisfies
\begin{align}
V^{+}(x;b_{1})=V^{-}(x;b_{1}+2d_{1})+b_{1}+d_{1},
\end{align}
which is consistent with the formula for $b_{1}^{(2)}$ in (\ref{eq:C14b1}), and thus
has two-step SI. The first and second intermediate potentials (\ref{eq:C14i1}) and
(\ref{eq:C14i2}) in this two-step SI case read respectively as
\begin{align}
V^{\rmi1}(x;b_{1}):=&\;V^{\rmi1}\Bigl(x;b_{1},\frac{a_{2}+d_{1}}{3a_{3}}b_{1}\Bigr)
 \biggr|_{a_{1}=\frac{(a_{2})^{2}-(d_{1})^{2}}{3a_{3}},\ a_{0}=\frac{(a_{2}+d_{1})^{2}
 (a_{2}-2d_{1})}{27(a_{3})^{2}}}\notag\\
=&\;\frac{15a_{3}}{16}z+\frac{5a_{2}}{16}+\frac{3(2b_{1}+d_{1})(2b_{1}-d_{1})}{
 16(3a_{3}z+a_{2}-2d_{1})}\biggr|_{z=z(x)},
\label{eq:C14i1''}
\end{align}
and
\begin{align}
V^{\rmi2}(x;b_{1}):=&\;V^{\rmi2}\Bigl(x;b_{1},\frac{a_{2}+d_{1}}{3a_{3}}b_{1}\Bigr)
 \biggr|_{a_{1}=\frac{(a_{2})^{2}-(d_{1})^{2}}{3a_{3}},\ a_{0}=\frac{(a_{2}+d_{1})^{2}
 (a_{2}-2d_{1})}{27(a_{3})^{2}}}\notag\\
=&\;-\frac{a_{3}}{16}z-\frac{a_{2}+16d_{1}}{48}+\frac{3(2b_{1}+d_{1})(2b_{1}-d_{1})}{
 16(3a_{3}z+a_{2}-2d_{1})}\biggr|_{z=z(x)}.
\label{eq:C14i2''}
\end{align}
Although both of them have the same functional dependence on the variable $z$
as $V^{\pm}$, neither has ordinary SI with $V^{\pm}$. Hence, the system has
irreducible two-step SI. The equation (\ref{eq:zq}) with $A(z)$ given in (\ref{eq:C14A''})
is integrated as
\begin{align}
z+\frac{a_{2}+d_{1}}{3a_{3}}=\left\{
\begin{array}{ll}
 \displaystyle{\frac{d_{1}}{a_{3}}\,\mathrm{sech}^{2}\,\sqrt{-\frac{d_{1}}{2}}}\,x\qquad
  &\text{for}\qquad d_{1}\neq 0,\\
 \displaystyle{\frac{2}{\sqrt{a_{3}}\,x^{2}}}&\text{for}\qquad d_{1}=0,
\end{array}\right.
\label{eq:C14z}
\end{align}
Substituting (\ref{eq:C14z}) into (\ref{eq:C14V''}), (\ref{eq:C14i1''}), and (\ref{eq:C14i2''}),
we finally obtain the potentials in the $x$-space.\\

\noindent
\textbf{Case 1-5:}  $a_{4}\neq0$

In the last, we shall consider the case of a fourth degree $A(z)$.
In this case, the pair of potentials (\ref{eq:V+-}) are given by
\begin{align}
\lefteqn{
V^{\pm}(x;b_{1},b_{0})=\frac{1}{4A(z)}\left[\left(3a_{3}a_{2}-6a_{4}a_{1}
 -\frac{3(a_{3})^{3}}{4a_{4}}\mp2a_{3}b_{1}\pm8a_{4}b_{0}\right)z^{3}\right.}\notag\\
&\;+\left(3(a_{2})^{2}-12a_{4}a_{0}-\frac{3a_{3}a_{1}}{2}-\frac{3(a_{3})^{2}a_{2}}{4a_{4}}
 \mp 4 a_{2}b_{1}\pm6a_{3}b_{0}+(b_{1})^{2}\right)z^{2}\notag\\
&\;+\left(3a_{2}a_{1}-6a_{3}a_{0}-\frac{3(a_{3})^{2}a_{1}}{4a_{4}}\mp6a_{1}b_{1}
 \pm4a_{2}b_{0}+2b_{1}b_{0}\right)z\notag\\
&\;\left.+\frac{3(a_{1})^{2}}{4}-\frac{3(a_{3})^{2}a_{0}}{4a_{4}}\mp8a_{0}b_{1}\pm2a_{1}b_{0}
 +(b_{0})^{2}\right]-\frac{a_{2}}{2}+\frac{3(a_{3})^{2}}{16a_{4}}\pm b_{1}\biggr|_{z=z(x)}.
\label{eq:C15V}
\end{align}
The first intermediate potential (\ref{eq:Vi1}) is calculated as
\begin{align}
V^{\rmi1}(x;b_{1},b_{0})=&\;
 \frac{1}{16A(z)}\left[\left(8a_{4}a_{1}-4a_{3}a_{2}+\frac{(a_{3})^{3}}{a_{4}}\right)z^{3}
 +\biggl(4(b_{1})^{2}+16a_{4}a_{0}+2a_{3}a_{1}\right.\notag\\
&\;-4(a_{2})^{2}+\frac{(a_{3})^{2}a_{2}}{a_{4}}\biggr)z^{2}
+\left(8b_{1}b_{0}+8a_{3}a_{0}-4a_{2}a_{1}+\frac{(a_{3})^{2}a_{1}}{a_{4}}\right)z\notag\\
&\;+\left.4(b_{0})^{2}-(a_{1})^{2}+\frac{(a_{3})^{2}a_{0}}{a_{4}}\right]+2a_{4}z^{2}+a_{3}z
 +\frac{a_{2}}{2}-\frac{(a_{3})^{2}}{16a_{4}}\biggr|_{z=z(x)},
\label{eq:C15i1}
\end{align}
while the second intermediate potential (\ref{eq:Vi2}) which exists only
when $b_{1}\neq0$ is as
\begin{align}
V^{\rmi2}(x;b_{1},b_{0})=&\;V^{\rmi1}(x;b_{1},b_{0})-2a_{4}z^{2}-a_{3}z
 -\frac{a_{3}b_{1}b_{0}-2a_{4}(b_{0})^{2}}{(b_{1})^{2}}\notag\\
&\;+\frac{a_{1}(b_{1})^{3}-2a_{2}(b_{1})^{2}b_{0}+3a_{3}b_{1}(b_{0})^{2}-4a_{4}(b_{0})^{3}}{(b_{1})^{2}
 (b_{1}z+b_{0})}\notag\\
&\;+2\frac{a_{0}(b_{1})^{4}-a_{1}(b_{1})^{3}b_{0}+a_{2}(b_{1})^{2}(b_{0})^{2}-a_{3}b_{1}(b_{0})^{3}
 +a_{4}(b_{0})^{4}}{(b_{1})^{2}(b_{1}z+b_{0})^{2}}\biggr|_{z=z(x)}.
\label{eq:C15i2}
\end{align}
Although $V^{+}$ and $V^{-}$ have the same functional dependence on the variable
$z$, it does not necessarily mean that there is a transformation of the parameters $b_{1}$
and $b_{0}$ which lead to non-reflective two-step SI. In fact, they satisfy the condition for
two-step SI (\ref{eq:tssi}) if and only if
\begin{align}
\left(\begin{array}{cc} a_{3} & -4a_{4}\\ b_{1}^{-}+4a_{2}\ & -6a_{3}\\
 b_{0}^{-}+6a_{1} & b_{1}^{-}-4a_{2}\\ 8a_{0} & b_{0}^{-}-2a_{1}\end{array}\right)\left(
 \begin{array}{ll} b_{1}^{+}\\ b_{0}^{+}\end{array}\right)=0,
\label{eq:C150}
\end{align}
where $b_{i}^{\pm}=b_{i}^{(2)}\pm b_{i}^{(0)}$. The trivial solution $b_{1}^{+}=b_{0}^{+}=0$
just produces reflective two-step SI (\ref{eq:refts}). To have a non-trivial solution,
the parameters must satisfy
\begin{align}
&2a_{4}(b_{1}^{-}+4a_{2})=3(a_{3})^{2},
\label{eq:C151}\\
&a_{3}(b_{1}^{-}-4a_{2})=-4a_{4}(b_{0}^{-}+6a_{1}),
\label{eq:C152}\\
&a_{3}(b_{0}^{-}-2a_{1})=-32a_{4}a_{0}.
\label{eq:C153}
\end{align}
We first consider the particular case when $a_{3}=0$. In the latter case, the set of
equations (\ref{eq:C151})--(\ref{eq:C153}) reduces by the assumption $a_{4}\neq0$ to
\begin{align}
b_{0}^{+}=b_{1}^{-}+4a_{2}=b_{0}^{-}+6a_{1}=a_{0}=0.
\end{align}
Hence, we have
\begin{align}
b_{1}^{(2)}=b_{1}^{(0)}-4a_{2},\quad b_{0}^{(2)}=-b_{0}^{(0)}=-3a_{1},\quad
 A(z)=a_{4}z^{4}+a_{2}z^{2}+a_{1}z.
\label{eq:C154}
\end{align}
In this case, the pair of potentials (\ref{eq:C15V}) reads as
\begin{align}
V^{\pm}(x;b_{1},b_{0})=&\;\frac{1}{4A(z)}\biggl[-2a_{4}(3a_{1}\mp 4b_{0})z^{3}
 +(3a_{2}\mp b_{1})(a_{2}\mp b_{1})z^{2}\notag\\
&\;+(3a_{2}a_{1}\mp 6a_{1}b_{1}\pm 4a_{2}b_{0}+2b_{1}b_{0})z+\frac{3(a_{1})^{2}}{4}\notag\\
&\;\mp 8a_{0}b_{1}\pm 2a_{1}b_{0}+(b_{0})^{2}\biggr]-\frac{a_{2}}{2}\pm b_{1}.
\label{eq:C15V'}
\end{align}
We can easily check that the latter pair satisfies
\begin{align}
V^{+}(x;b_{1},b_{0})A(z)=&\;V^{-}(x;b_{1}-4a_{2},-b_{0})A(z)\notag\\
&\;+(b_{1}-2a_{2})\left[2a_{4}z^{4}+2a_{2}z^{2}+(b_{0}-a_{1})z\right],
\end{align}
and thus has two-step SI if and only if $b_{0}=3a_{1}$, as has been indicated by
the second equality in (\ref{eq:C154}). Hence, the present system has conditional
two-step SI.

The first and second intermediate potentials (\ref{eq:C15i1}) and (\ref{eq:C15i2})
in this case of conditional two-step SI read respectively as
\begin{align}
V^{\rmi1}(x;b_{1},b_{0})=&\;\frac{1}{4A(z)}\biggl[2a_{4}a_{1}z^{3}+\left((b_{1})^{2}
 -(a_{2})^{2}\right)z^{2}+(2b_{1}b_{0}-a_{2}a_{1})z\notag\\
&\;+(b_{0})^{2}-\frac{(a_{1})^{2}}{4}\biggr]+2a_{4}z^{2}+\frac{a_{2}}{2}\biggr|_{z=z(x)},
\label{eq:C15i1'}
\end{align}
and
\begin{align}
V^{\rmi2}(x;b_{1},b_{0})=&\;V^{\rmi1}(x;b_{1},b_{0})-2a_{4}z^{2}+\frac{2a_{4}(b_{0})^{2}
 }{(b_{1})^{2}}+\frac{a_{1}(b_{1})^{3}-2a_{2}(b_{1})^{2}b_{0}-4a_{4}(b_{0})^{3}}{(b_{1})^{2}
 (b_{1}z+b_{0})}\notag\\
&\;-2b_{0}\frac{a_{1}(b_{1})^{3}-a_{2}(b_{1})^{2}b_{0}-a_{4}(b_{0})^{3}}{(b_{1})^{2}
 (b_{1}z+b_{0})^{2}}\biggr|_{z=z(x)}.
\label{eq:C15i2'}
\end{align}
Both of them do not have ordinary SI with $V^{\pm}$ given in (\ref{eq:C15V'})
irrespective of whether $b_{0}=3a_{1}$ or not.
Hence, the conditional two-step SI of the system (\ref{eq:C15V'}) is irreducible.

The general solution to equation (\ref{eq:zq}) with $A(z)$ given in (\ref{eq:C154})
is expressed in terms of elliptic functions when $a_{1}\neq0$. The potentials are
thus of deformed Lam\'{e} or Heun type. We omit its involved general form here. When
$a_{1}=0$, the system automatically has two-step SI since $b_{0}^{(2)}=b_{0}^{(0)}=0$
from the second equality in (\ref{eq:C154}). In addition, the form of $A(z)$
reduces to the one given in (\ref{eq:C15A''}) with $a_{3}=0$, and the solution
to (\ref{eq:zq}) is given by (\ref{eq:C15z}). Substituting (\ref{eq:C15z}) into
(\ref{eq:C15V'}), (\ref{eq:C15i1'}), and (\ref{eq:C15i2'}), we finally obtain
the potentials in the $x$-space.\\

When $a_{3}\neq0$, we immediately obtain from (\ref{eq:C151}) and (\ref{eq:C153})
\begin{align}
b_{1}^{-}=-4a_{2}+\frac{3(a_{3})^{2}}{2a_{4}},\qquad
 b_{0}^{-}=-\frac{32a_{4}a_{0}}{a_{3}}+2a_{1}.
\label{eq:C15b1}
\end{align}
Substituting them into (\ref{eq:C152}), we have the following constraint
\begin{align}
64(a_{4})^{2}(4a_{4}a_{0}-a_{3}a_{1})=-(a_{3})^{2}\left[16a_{4}a_{2}-3(a_{3})^{2}\right].
\label{eq:C15c1}
\end{align}
On the other hand, the homogeneous equations (\ref{eq:C150}) under the conditions
(\ref{eq:C151})--(\ref{eq:C153}) lead to a single equation $a_{3}b_{1}^{+}=4a_{4}b_{0}^{+}$.
Eliminating $b_{1}^{(2)}$, $b_{0}^{(2)}$, and $a_{0}$ in it by using (\ref{eq:C15b1}) and
(\ref{eq:C15c1}), we obtain
\begin{align}
a_{3}b_{1}^{(0)}-4a_{4}b_{0}^{(0)}+12a_{4}a_{1}-6a_{3}a_{2}+\frac{3(a_{3})^{3}}{2a_{4}}=0.
\end{align}
This relation means that $b_{1}$ and $b_{0}$ cannot be independent parameters.
It is evident that it must hold not only between $b_{1}^{(0)}$ and $b_{0}^{(0)}$ but also
between $b_{1}^{(2)}$ and $b_{0}^{(2)}$ for the consistency. From the latter requirement,
we have another constraint:
\begin{align}
8a_{4}a_{1}-4a_{3}a_{2}+\frac{(a_{3})^{3}}{a_{4}}=0.
\label{eq:C15c3}
\end{align}
Combining (\ref{eq:C15c1})--(\ref{eq:C15c3}), we finally obtain
\begin{align}
\begin{split}
&b_{0}=\frac{a_{3}}{4a_{4}}b_{1},\qquad b_{1}^{(2)}=b_{1}^{(0)}-4a_{2}+\frac{3(a_{3})^{2}}{2a_{4}},\\
&a_{1}=\frac{a_{3}a_{2}}{2a_{4}}-\frac{(a_{3})^{3}}{8(a_{4})^{2}},\qquad
 a_{0}=\frac{(a_{3})^{2}a_{2}}{16(a_{4})^{2}}-\frac{5(a_{3})^{4}}{256(a_{4})^{3}}.
\end{split}
\label{eq:C15f}
\end{align}
Under the latter parameter relations, the function $A(z)$ is factorized as
\begin{align}
A(z)=a_{4}\left(z+\frac{a_{3}}{4a_{4}}\right)^{2}\left(z^{2}+\frac{a_{3}}{2a_{4}}z
 +\frac{a_{2}}{a_{4}}-\frac{5(a_{3})^{2}}{16(a_{4})^{2}}\right),
\label{eq:C15A''}
\end{align}
and the pair of potentials (\ref{eq:C15V}) reads as
\begin{align}
V^{\pm}(x;b_{1}):=&\;V^{\pm}\Bigl(x;b_{1},\frac{a_{3}}{4a_{4}}b_{1}\Bigr)\biggr|_{a_{1}=
 \frac{a_{3}a_{2}}{2a_{4}}-\frac{(a_{3})^{3}}{8(a_{4})^{2}},\ 
 a_{0}=\frac{(a_{3})^{2}a_{2}}{16(a_{4})^{2}}-\frac{5(a_{3})^{4}}{256(a_{4})^{3}}}\notag\\
=&\;\frac{4a_{4}}{16(a_{4})^{2}z^{2}+8a_{4}a_{3}z+16a_{4}a_{2}-5(a_{3})^{2}}\notag\\
&\;\times\left(
 a_{2}-\frac{3(a_{3})^{2}}{8a_{4}}\mp b_{1}\right)\left(3a_{2}-\frac{9(a_{3})^{2}}{8a_{4}}
 \mp b_{1}\right)-\frac{a_{2}}{2}+\frac{3(a_{3})^{2}}{16a_{4}}\pm b_{1}\biggr|_{z=z(x)}.
\label{eq:C15V''}
\end{align}
We now easily check that the latter pair of potentials satisfies
\begin{align}
V^{+}(x;b_{1})=V^{-}\biggl(x;b_{1}-4a_{2}+\frac{3(a_{3})^{2}}{2a_{4}}\biggr)
 +2\left[b_{1}-2a_{2}+\frac{3(a_{3})^{2}}{4a_{4}}\right],
\end{align}
which is consistent with the formula for $b_{1}^{(2)}$ in (\ref{eq:C15f}), and thus
has two-step SI. The first and second intermediate potentials (\ref{eq:C15i1}) and
(\ref{eq:C15i2}) in this two-step SI case read respectively as
\begin{align}
V^{\rmi1}(x;b_{1}):=&\;V^{\rmi1}\Bigl(x;b_{1},\frac{a_{3}}{4a_{4}}b_{1}\Bigr)\biggr|_{a_{1}=
 \frac{a_{3}a_{2}}{2a_{4}}-\frac{(a_{3})^{3}}{8(a_{4})^{2}},\ 
 a_{0}=\frac{(a_{3})^{2}a_{2}}{16(a_{4})^{2}}-\frac{5(a_{3})^{4}}{256(a_{4})^{3}}}\notag\\
=&\;\frac{4a_{4}}{16(a_{4})^{2}z^{2}+8a_{4}a_{3}z+16a_{4}a_{2}-5(a_{3})^{2}}\notag\\
&\;\times\left[(b_{1})^{2}-\left(a_{2}-\frac{3(a_{3})^{2}}{8a_{4}}\right)^{2}\right]
 +2a_{4}z^{2}+a_{3}z+\frac{a_{2}}{2}-\frac{(a_{3})^{2}}{16a_{4}}\biggr|_{z=z(x)},
\label{eq:C15i1''}
\end{align}
and
\begin{align}
V^{\rmi2}(x;b_{1}):=&\;V^{\rmi2}\Bigl(x;b_{1},\frac{a_{3}}{4a_{4}}b_{1}\Bigr)\biggr|_{a_{1}=
 \frac{a_{3}a_{2}}{2a_{4}}-\frac{(a_{3})^{3}}{8(a_{4})^{2}},\ 
 a_{0}=\frac{(a_{3})^{2}a_{2}}{16(a_{4})^{2}}-\frac{5(a_{3})^{4}}{256(a_{4})^{3}}}\notag\\
=&\;\frac{4a_{4}}{16(a_{4})^{2}z^{2}+8a_{4}a_{3}z+16a_{4}a_{2}-5(a_{3})^{2}}\notag\\
&\;\times\left[(b_{1})^{2}-\left(a_{2}-\frac{3(a_{3})^{2}}{8a_{4}}\right)^{2}\right]
 +\frac{a_{2}}{2}-\frac{3(a_{3})^{2}}{16a_{4}}\biggr|_{z=z(x)}.
\label{eq:C15i2''}
\end{align}
Hence, the second intermediate potential $V^{\rmi2}$ together with $V^{\pm}$
constitutes a sequence of ordinary SI as
\begin{align}
\begin{split}
V^{\rmi2}(x;b_{1})&=V^{-}\Bigl(x;b_{1}-2a_{2}+\frac{3(a_{3})^{2}}{4a_{4}}\Bigr)
 +b_{1}-a_{2}+\frac{3(a_{3})^{2}}{8a_{4}},\\
V^{+}(x;b_{1})&=V^{\rmi2}\Bigl(x;b_{1}-2a_{2}+\frac{3(a_{3})^{2}}{4a_{4}}\Bigr)
 +b_{1}-a_{2}+\frac{3(a_{3})^{2}}{8a_{4}}.
\end{split}
\end{align}
Therefore, the system has reducible two-step SI. The equation (\ref{eq:zq}) with $A(z)$
given in (\ref{eq:C15A''}) is integrated as
\begin{align}
z+\frac{a_{3}}{4a_{4}}=\left\{
\begin{array}{ll}
 \displaystyle{\frac{4\sqrt{c}\,\rme^{\sqrt{2a_{4}c}\,x}}{\rme^{2\sqrt{2a_{4}c}\,x}-4}}\qquad
  &\text{for}\qquad c\neq 0,\\
 \displaystyle{-\frac{1}{\sqrt{2a_{4}}\,x}}&\text{for}\qquad c=0,
\end{array}\right.
\label{eq:C15z}
\end{align}
where
\begin{align}
c=\frac{a_{2}}{a_{4}}-\frac{3(a_{3})^{2}}{8(a_{4})^{2}}.
\end{align}
Substituting (\ref{eq:C15z}) into (\ref{eq:C15V''}), (\ref{eq:C15i1''}), and (\ref{eq:C15i2''}),
we finally obtain the potentials in the $x$-space.\\

To summarize, we have found that a type A $2$-fold SUSY system with an intermediate
Hamiltonian resulting from a polynomial $A(z)$ of at most fourth degree can possess
non-reflective two-step SI, always when the degree is less than or equal to two, and
under certain conditions when it is three or four. 
Another remarkable result is that all the type A 2-fold SUSY systems which have
two-step SI as well in Cases 1-1 to 1-3 turn to have also ordinary SI, that is, they have
reducible two-step SI, and they are all well-known SI potentials. On the other hand,
all the other two-step SI potentials in Cases 1-4 and 1-5 are, to the best of our
knowledge, new and except for the last model in Case 1-5 they do not possess ordinary
SI, that is, they have irreducible two-step SI.
In the next section, we shall examine more general cases where the function $A(z)$ is
not given by a polynomial.

\section{Non-polynomial $A(z)$}
\label{sec:nt2SI}

To investigate the possibility of other irreducible two-step SI, let us coming back to
the form of potentials (\ref{eq:V+-}). In general, the function $A(z)$ can depend
on the parameters $b_{i}$. In this case, the change of variables $z=z(x)$ also depends
on them via (\ref{eq:zq}), and analysis of the two-step SI condition (\ref{eq:tssi})
with the pair of potentials (\ref{eq:V+-}) gets quite involved. In this work, we shall
restrict our analysis to the case where $A(z)$, and thus $z=z(x)$ as well, does not
depend on the parameters $b_{i}$. Under this restriction, we easily see from
(\ref{eq:V+-}) that the two-step SI condition (\ref{eq:tssi}) holds if and only if
the function $A(z)$ satisfies
\begin{align}
&\frac{1}{4A(z)}\Bigl[(b_{1}^{(0)}z+b_{0}^{(0)})^{2}+2\bigl((b_{1}^{(0)}z+b_{0}^{(0)})
 A'(z)-2b_{1}^{(0)}A(z)\bigr)\Bigr]\notag\\
&=\frac{1}{4A(z)}\Bigl[(b_{1}^{(2)}z+b_{0}^{(2)})^{2}-2\bigl((b_{1}^{(2)}z
 +b_{0}^{(2)})A'(z)-2b_{1}^{(2)}A(z)\bigr)\Bigr]+\tR_{2}(b_{1}^{(0)},b_{0}^{(0)}).
\end{align}
The latter condition is identical to the first-order linear differential equation
\begin{align}
(b_{1}^{+}z+b_{0}^{+})(b_{1}^{-}z+b_{0}^{-})-2(b_{1}^{+}z+b_{0}^{+})A'(z)+4\bar{R}A(z)=0,
\label{eq:cAz1}
\end{align}
where $b_{i}^{\pm}=b_{i}^{(2)}\pm b_{i}^{(0)}$ and $\bar{R}=\tR_{2}+b_{1}^{+}$. It is
apparent that its most general solution $A(z)$ depends on the parameters $b_{i}^{(0)}$.
Hence, to obtain a solution without such dependence, we must impose the following
additional condition:
\begin{align}
\frac{\del A(z)}{\del b_{i}^{(0)}}=0.
\label{eq:cAz2}
\end{align}
As we will show in what follows, the latter condition completely determines the
admissible form of the function $A(z)$ and the parameters $b_{i}^{(2)}$ and $\tR_{2}$.

To solve the differential equation (\ref{eq:cAz1}), we first note that we can assume
without any loss of generality that either $b_{1}^{+}$ or $b_{0}^{+}$ is non-zero.
Otherwise, we have
\begin{align}
b_{1}^{(2)}=-b_{1}^{(0)},\qquad b_{0}^{(2)}=-b_{0}^{(0)},\qquad \tR_{2}=0,
\end{align}
which exactly leads to the two-step SI of the type (\ref{eq:refts}). Thus, $b_{1}^{+}z
+b_{0}^{+}$ is not identically zero, and (\ref{eq:cAz1}) is integrated as
\begin{align}
2A(z)=\exp\left(\int\rmd z\,\frac{2\bar{R}}{b_{1}^{+}z+b_{0}^{+}}\right)
 \int\rmd z\,(b_{1}^{-}z+b_{0}^{-})\exp\left(-\int^{z}\rmd z'\,\frac{2\bar{R}}{
 b_{1}^{+}z'+b_{0}^{+}}\right).
\label{eq:Az1}
\end{align}
It requires separate treatments according to whether $b_{1}^{+}$ is zero or not.\\

\noindent
\textbf{Case 2:} $b_{1}^{+}\neq0$

Let us first study the case of non-zero $b_{1}^{+}$. In this case, equation (\ref{eq:Az1})
reads as
\begin{align}
2A(z)&=(b_{1}^{+}z+b_{0}^{+})^{\mu}\int\rmd z\,(b_{1}^{-}z+b_{0}^{-})
 (b_{1}^{+}z+b_{0}^{+})^{-\mu}\notag\\
&=(b_{1}^{+}z+b_{0}^{+})^{\mu}\int\rmd z\,\left[\frac{b_{1}^{-}}{b_{1}^{+}}
 (b_{1}^{+}z+b_{0}^{+})^{1-\mu}+\frac{b_{1}^{+}b_{0}^{-}-b_{1}^{-}b_{0}^{+}
 }{b_{1}^{+}}(b_{1}^{+}z+b_{0}^{+})^{-\mu}\right],
\label{eq:Az2}
\end{align}
where $\mu=2\bar{R}/b_{1}^{+}$. Hence, we have the following three inequivalent cases.\\

\noindent
\textbf{Case 2-1:} $\mu\neq1,2$

In this case, we obtain from (\ref{eq:Az2})
\begin{align}
A(z)=a_{2}z^{2}+a_{1}z+a_{0}+c(b_{1}^{+}z+b_{0}^{+})^{\mu},
\end{align}
where $c$ is an integral constant and
\begin{align}
\begin{split}
a_{2}&=\frac{b_{1}^{-}}{2(2-\mu)},\qquad a_{1}=-\frac{\mu b_{1}^{-}b_{0}^{+}
 }{2(1-\mu)(2-\mu)b_{1}^{+}}+\frac{b_{0}^{-}}{2(1-\mu)},\\
a_{0}&=-\frac{b_{1}^{-}(b_{0}^{+})^{2}}{2(1-\mu)(2-\mu)(b_{1}^{+})^{2}}
 +\frac{b_{0}^{+}b_{0}^{-}}{2(1-\mu)b_{1}^{+}}.
\end{split}
\label{eq:para1}
\end{align}
The condition (\ref{eq:cAz2}) in this case is equivalent to the following set
of equations:
\begin{align}
\frac{\del a_{2}\:\:}{\del b_{i}^{(0)}}=\frac{\del a_{1}\:\:}{\del b_{i}^{(0)}}=
 \frac{\del a_{0}\:\:}{\del b_{i}^{(0)}}=b_{1}^{+}\frac{\del c\:\;\;}{\del b_{i}^{(0)}}+c\mu
 \frac{\del b_{1}^{+}\:\:}{\del b_{i}^{(0)}}=b_{0}^{+}\frac{\del c\:\;\;}{\del b_{i}^{(0)}}
 +c\mu\frac{\del b_{0}^{+}\:\:}{\del b_{i}^{(0)}}=\frac{\del\mu\:\;\;}{\del b_{i}^{(0)}}=0.
\label{eq:cond1}
\end{align}
We assume $c\neq0$ to avoid duplication of Case 1-3.
The last equality just means that $\mu$ does not depend on $b_{i}^{(0)}$. Substituting
the expression of the parameters (\ref{eq:para1}) into (\ref{eq:cond1}),
we see that the following equations must hold:
\begin{align}
\frac{\del b_{1}^{-}\:\:}{\del b_{i}^{(0)}}=0,\qquad\frac{\del b_{0}^{-}\:\:}{\del b_{i}^{(0)}}=0,
 \qquad b_{0}^{+}\frac{\del b_{1}^{+}\:\:}{\del b_{i}^{(0)}}
 -b_{1}^{+}\frac{\del b_{0}^{+}\:\:}{\del b_{i}^{(0)}}=0.
\label{eq:cond1'}
\end{align}
{}From the first and second equalities in (\ref{eq:cond1'}), we immediately have
\begin{align}
b_{1}^{(2)}=b_{1}^{(0)}+2d_{1},\qquad b_{0}^{(2)}=b_{0}^{(0)}+2d_{0},
\label{eq:cond1''}
\end{align}
where $d_{1}$ and $d_{0}$ are constants which do not depend on $b_{i}^{(0)}$.

Until now, we have not made any assumption on the relation between $b_{1}^{(0)}$ and
$b_{0}^{(0)}$, that is, they can be independent or dependent parameters. So, let us first
assume that they are independent. In this case, we shall first show that $b_{0}^{+}=0$.
For this purpose, we note that so long as $b_{0}^{+}=2b_{0}^{(0)}+2d_{0}\neq0$ together
with the assumption $b_{1}^{+}=2b_{1}^{(0)}+2d_{1}\neq0$
\begin{align}
\frac{\del b_{1}^{+}\:\:}{\del b_{1}^{(0)}}=\frac{\del b_{0}^{+}\:\:}{\del b_{0}^{(0)}}=2,\qquad
 \frac{\del b_{0}^{+}\:\:}{\del b_{1}^{(0)}}=\frac{\del b_{1}^{+}\:\:}{\del b_{0}^{(0)}}=0.
\end{align}
Then, the third equation in (\ref{eq:cond1'}) results in $b_{1}^{+}=b_{0}^{+}=0$, which
is obviously contradictory. Hence, we must have $b_{0}^{+}=0$, which together with
(\ref{eq:cond1''}) means
\begin{align}
b_{1}^{(2)}=b_{1}^{(0)}+2d_{1},\qquad b_{0}^{(2)}=-b_{0}^{(0)}=d_{0},
\label{eq:sol1}
\end{align}
and the third condition in (\ref{eq:cond1'}) is trivially satisfied. All the remaining
equations in (\ref{eq:cond1}) to be satisfied now read as
\begin{align}
\frac{\del c\:\;\;}{\del b_{1}^{(0)}}+\frac{2\mu}{b_{1}^{+}}c
 =\frac{\del c\:\;\;}{\del b_{0}^{(0)}}=0,
\end{align}
and their general solution is given by
\begin{align}
c=c_{0}/(b_{1}^{+})^{\mu},
\label{eq:sol1''}
\end{align}
where $c_{0}$ and $\mu$ are constants which do not depend on $b_{1}^{(0)}$.
Substituting (\ref{eq:sol1}) back into (\ref{eq:para1}), we have
\begin{align}
d_{1}=-(\mu-2)a_{2},\qquad d_{0}=-(\mu-1)a_{1},\qquad a_{0}=0.
\label{eq:sol1'}
\end{align}
With the aid of the solutions (\ref{eq:sol1}), (\ref{eq:sol1''}), and (\ref{eq:sol1'}),
we finally obtain the admissible form of the function $A(z)$ and the parameters
$b_{i}^{(2)}$ and $\tR_{2}$ as
\begin{align}
\begin{split}
&A(z)=a_{2}z^{2}+a_{1}z+c_{0}z^{\mu},\qquad b_{1}^{(2)}=b_{1}^{(0)}-2(\mu-2)a_{2},\\
&b_{0}^{(2)}=-b_{0}^{(0)}=-(\mu-1)a_{1},\qquad
 \tR_{2}=(\mu-2)\bigl[b_{1}^{(0)}-(\mu-2)a_{2}\bigr].
\end{split}
\label{eq:adm1}
\end{align}
The pair of potentials (\ref{eq:V+-}) in this case are given by
\begin{align}
\lefteqn{
16V^{\pm}(x;b_{1},b_{0})A(z(x))=4\bigl[(a_{2})^{2}+(b_{1})^{2}\bigr] z^{2}
 +4(a_{2}a_{1}\mp 2a_{1}b_{1}\pm 4a_{2}b_{0}+2b_{1}b_{0})z}\hspace{50pt}\notag\\
&\;+3(a_{1})^{2}\pm 8a_{1}b_{0}+4(b_{0})^{2}-4\left[(\mu^{2}-4\mu+2)a_{2}
 \mp 2(\mu-2)b_{1}\right]c_{0}z^{\mu}\notag\\
&\;-2\mu\left[(2\mu-5)a_{1}\mp 4b_{0}\right]c_{0}z^{\mu-1}
 -\mu(\mu-4)(c_{0})^{2}z^{2\mu-2}\bigr|_{z=z(x)}.
\label{eq:C21V}
\end{align}
We can easily check that the latter pair satisfies
\begin{align}
V^{+}(x;b_{1},b_{0})A(z)=&\;V^{-}(x;b_{1}-2(\mu-2)a_{2},-b_{0})A(z)\notag\\
&\;+[b_{1}-(\mu-2)a_{2}]\left[(\mu-2)a_{2}z^{2}+(b_{0}-a_{1})z+(\mu-2)c_{0}z^{\mu}\right],
\end{align}
and thus has two-step SI if and only if $b_{0}=(\mu-1)a_{1}$, as has been indicated
by the third equality in (\ref{eq:adm1}). Hence, the present system has conditional
two-step SI.

We note that $b_{1}\neq0$ in this case and the system always admits two intermediate
Hamiltonians. The first and second intermediate potentials (\ref{eq:Vi1}) and
(\ref{eq:Vi2}) are respectively calculated as
\begin{align}
\lefteqn{
16V^{\rmi1}(x;b_{1},b_{0})A(z(x))=4\bigl[(a_{2})^{2}+(b_{1})^{2}\bigr] z^{2}
 +4(a_{2}a_{1}+2b_{1}b_{0})z-(a_{1})^{2}+4(b_{0})^{2}}\hspace{30pt}\notag\\
&\;+4(\mu^{2}-2\mu+2)a_{2}c_{0}z^{\mu}+2\mu(2\mu-3)a_{1}c_{0}z^{\mu-1}
 +\mu(3\mu-4)(c_{0})^{2}z^{2\mu-2}\bigr|_{z=z(x)},
\label{eq:C21i1}
\end{align}
and
\begin{align}
V^{\rmi2}(x;b_{1},b_{0})=&\;V^{\rmi1}(x;b_{1},b_{0})+\frac{a_{1}b_{1}-2a_{2}b_{0}
 -(\mu-2)c_{0}b_{1}z^{\mu-1}}{b_{1}z+b_{0}}\notag\\\
&\;-2b_{0}\frac{a_{1}b_{1}-a_{2}b_{0}+c_{0}b_{1}z^{\mu-1}}{(b_{1}z+b_{0})^{2}}\biggr|_{z=z(x)}.
\label{eq:C21i2}
\end{align}
Both of them do not have ordinary SI with $V^{\pm}$ given in (\ref{eq:C21V})
irrespective of whether $b_{0}=(\mu-1)a_{1}$ or not.
Hence, the conditional two-step SI of the system (\ref{eq:C21V}) is irreducible.
We note that the present system (\ref{eq:adm1})--(\ref{eq:C21i2}) reduces to the one
(\ref{eq:C144})--(\ref{eq:C14i2'}) in Case 1-4 when we put $\mu=3$ with $c_{0}=a_{3}$
and to the one (\ref{eq:C154})--(\ref{eq:C15i2'}) in Case 1-5 when we put $\mu=4$
with $c_{0}=a_{4}$.

The equation (\ref{eq:zq}) with $A(z)$ given in (\ref{eq:adm1}) cannot be integrated
analytically unless $a_{1}=0$. When $a_{1}=0$, the system automatically has two-step
SI since $b_{0}^{(2)}=b_{0}^{(0)}=0$ from the third equality in (\ref{eq:adm1}). In
addition, the form of $A(z)$ reduces to the one given in (\ref{eq:adm1'}) with $a_{1}=0$,
and the solution to (\ref{eq:zq}) is given by (\ref{eq:C21z}). Substituting (\ref{eq:C21z})
into (\ref{eq:C21V}), (\ref{eq:C21i1}), and (\ref{eq:C21i2}), we finally obtain the potentials
in the $x$-space.\\

In the next, let us consider the case when $b_{1}^{(0)}$ and $b_{0}^{(0)}$ are dependent.
In this case, we can assume that $b_{0}^{(0)}$ is a function of $b_{1}^{(0)}$. The set of
conditions (\ref{eq:cond1}), and (\ref{eq:cond1'}) as well, should be considered only for
$i=1$. The solution (\ref{eq:cond1''}) to the first and second equations in (\ref{eq:cond1'})
is still valid, but the one to the third one is $b_{0}^{+}=z_{1}b_{1}^{+}$ where $z_{1}$
is a constant which does not depend on $b_{1}^{(0)}$. Substituting the latter into
(\ref{eq:cond1''}) and requiring that the functional dependence of $b_{0}^{(2)}$ on
$b_{1}^{(2)}$ must be the same as that of $b_{0}^{(0)}$ on $b_{1}^{(0)}$, we obtain
\begin{align}
b_{0}^{(0)}=z_{1}b_{1}^{(0)},\qquad d_{0}=z_{1}d_{1}.
\label{eq:dsol1}
\end{align}
All the remaining equations in (\ref{eq:cond1}) are satisfied with the solution
(\ref{eq:sol1''}). Substituting (\ref{eq:cond1''}) and (\ref{eq:dsol1}) back into
(\ref{eq:para1}), we find that $a_{2}\neq0$ is necessary for the non-triviality of
the system and obtain
\begin{align}
b_{1}^{(2)}=b_{1}^{(0)}-2(\mu-2)a_{2},\qquad 2a_{2}z_{1}=a_{1},\qquad 4a_{2}a_{0}=(a_{1})^{2}.
\label{eq:dsol2}
\end{align}
With the aid of the solutions (\ref{eq:cond1''}), (\ref{eq:dsol1}), and (\ref{eq:dsol2}),
we finally obtain the admissible form of the function $A(z)$ and the parameters
$b_{1}^{(2)}$ and $\tR_{2}$ as
\begin{align}
\begin{split}
&A(z)=a_{2}\left(z+\frac{a_{1}}{2a_{2}}\right)^{2}+c_{0}\left(z+\frac{a_{1}}{2a_{2}}
 \right)^{\mu},\qquad b_{1}^{(2)}=b_{1}^{(0)}-2(\mu-2)a_{2},\\
&b_{0}^{(2)}=\frac{a_{1}}{2a_{2}}b_{1}^{(2)},\qquad b_{0}^{(0)}=\frac{a_{1}}{2a_{2}}
 b_{1}^{(0)},\qquad \tR_{2}=(\mu-2)\bigl[b_{1}^{(0)}-(\mu-2)a_{2}\bigr].
\end{split}
\label{eq:adm1'}
\end{align}
The pair of potentials (\ref{eq:V+-}) in this case with the definition $V(x;b_{1}):=
V(x;b_{1},a_{1}b_{1}/(2a_{2}))$ are calculated as
\begin{align}
\lefteqn{
16V^{\pm}(x;b_{1})A(z(x))=4\bigl[(a_{2})^{2}+(b_{1})^{2}\bigr]\left(z+\frac{a_{1}}{2a_{2}}
 \right)^{2}-4\bigl[(\mu^{2}-4\mu+2)a_{2}}\hspace{50pt}\notag\\
&\;\mp2(\mu-2)b_{1}\bigr]c_{0}\left(z+\frac{a_{1}}{2a_{2}}\right)^{\mu}
 -\mu(\mu-4)(c_{0})^{2}\left(z+\frac{a_{1}}{2a_{2}}\right)^{2\mu-2}\biggr|_{z=z(x)}.
\label{eq:C21V'}
\end{align}
We can easily check that the latter pair satisfies
\begin{align}
V^{+}(x;b_{1})=V^{-}(x;b_{1}-2(\mu-2)a_{2})+(\mu-2)[b_{1}-(\mu-2)a_{2}],
\end{align}
which is consistent with the formulas for $b_{1}^{(2)}$ and $\tR_{2}$ in (\ref{eq:adm1'}),
and thus has two-step SI. The first and second intermediate potentials (\ref{eq:Vi1})
and (\ref{eq:Vi2}), which always exist thanks to $b_{1}\neq0$, are respectively given by
\begin{align}
16V^{\rmi1}(x;b_{1})A(z)=&\;4\bigl[(a_{2})^{2}+(b_{1})^{2}\bigr]\left(z+\frac{a_{1}}{2a_{2}}
 \right)^{2}+4(\mu^{2}-2\mu+2)a_{2}c_{0}\left(z+\frac{a_{1}}{2a_{2}}\right)^{\mu}\notag\\
&\;+\mu(3\mu-4)(c_{0})^{2}\left(z+\frac{a_{1}}{2a_{2}}\right)^{2\mu-2},
\label{eq:C21i1'}
\end{align}
and
\begin{align}
V^{\rmi2}(x;b_{1})=V^{\rmi1}(x;b_{1})-(\mu-2)c_{0}\left(z+\frac{a_{1}}{2a_{2}}
 \right)^{\mu-2}.
\label{eq:C21i2'}
\end{align}
Both of them in general do not have ordinary SI with $V^{\pm}$ given in (\ref{eq:C21V'}).
Hence, the present system has irreducible two-step SI in general. It is worth noting,
however, that the system (\ref{eq:C21V'})--(\ref{eq:C21i2'}) reduces to the \emph{reducible}
two-step SI system (\ref{eq:C15V''})--(\ref{eq:C15i2''}) in Case 1-5 when $\mu=4$.
Indeed, we can check that with the following substitution
\begin{align}
c_{0}=\bar{a}_{4},\qquad a_{2}=\bar{a}_{2}-\frac{3(\bar{a}_{3})^{2}}{8\bar{a}_{4}},\qquad
 a_{1}=\frac{\bar{a}_{3}\bar{a}_{2}}{2\bar{a}_{4}}-\frac{3(\bar{a}_{3})^{3}}{16(\bar{a}_{4})^{2}},
\end{align}
the system (\ref{eq:C21V'})--(\ref{eq:C21i2'}) coincides with (\ref{eq:C15V''})--(\ref{eq:C15i2''})
with $a_{i}$ replaced by $\bar{a}_{i}$ when $\mu=4$. On the other hand, when $\mu=3$,
the system (\ref{eq:C21V'})--(\ref{eq:C21i2'}) reduces to the one (\ref{eq:C14V''})--(\ref{eq:C14i2''})
in Case 1-4 by the following substitution
\begin{align}
c_{0}=\bar{a}_{3},\qquad a_{2}=-\bar{d}_{1},\qquad a_{1}=-\frac{2\bar{d}_{1}
 (\bar{a}_{2}+\bar{d}_{1})}{3\bar{a}_{3}},
\end{align}
keeping the irreducibility intact. Therefore, the reducibility of two-step SI
in the system (\ref{eq:C15V''})--(\ref{eq:C15i2''}) in Case 1-5 is rather exceptional.

The equation (\ref{eq:zq}) with $A(z)$ given in (\ref{eq:adm1'}) is integrated as
\begin{align}
\left(z+\frac{a_{1}}{2a_{2}}\right)^{2-\mu}=\frac{c_{0}}{a_{2}}\tanh^{2}(\mu-2)
 \sqrt{\frac{a_{2}}{2}}x.
\label{eq:C21z}
\end{align}
Substituting (\ref{eq:C21z}) into (\ref{eq:C21V'}), (\ref{eq:C21i1'}), and (\ref{eq:C21i2'}),
we finally obtain the potentials in the $x$-space.\\

\noindent
\textbf{Case 2-2:} $\mu=1$

In this case, we obtain from (\ref{eq:Az2})
\begin{align}
A(z)=a_{2}z^{2}+a_{1}z+a_{0}+\bar{c}(b_{1}^{+}z+b_{0}^{+})\ln|b_{1}^{+}z+b_{0}^{+}|,
\end{align}
where
\begin{align}
a_{2}=\frac{b_{1}^{-}}{2},\quad a_{1}=\frac{b_{1}^{-}b_{0}^{+}}{2b_{1}^{+}}+cb_{1}^{+},
 \quad a_{0}=cb_{0}^{+},\quad \bar{c}=\frac{b_{1}^{+}b_{0}^{-}-b_{1}^{-}b_{0}^{+}
 }{2(b_{1}^{+})^{2}},
\label{eq:para2}
\end{align}
and $c$ is an integral constant. The condition (\ref{eq:cAz2}) in this case is
equivalent to the following set of equations:
\begin{align}
\frac{\del a_{2}\:\:}{\del b_{i}^{(0)}}=\frac{\del a_{1}\:\:}{\del b_{i}^{(0)}}+\bar{c}
 \frac{\del b_{1}^{+}\:\:}{\del b_{i}^{(0)}}=\frac{\del a_{0}\:\:}{\del b_{i}^{(0)}}+\bar{c}
 \frac{\del b_{0}^{+}\:\:}{\del b_{i}^{(0)}}=\frac{\del (\bar{c}b_{1}^{+})}{\del b_{i}^{(0)}}
 =\frac{\del(\bar{c}b_{0}^{+})}{\del b_{i}^{(0)}}=0.
\label{eq:cond2}
\end{align}
We assume $\bar{c}\neq0$ to avoid duplication of Case 1-3.
Substituting the expression of the parameters (\ref{eq:para2}) into (\ref{eq:cond2}),
we see that the condition (\ref{eq:cond1'}) must hold also in the present case. So, let
us first investigate the case when $b_{1}^{(0)}$ and $b_{0}^{(0)}$ are independent. In
this case, by following the same argument below (\ref{eq:cond1'}), we arrive at the same
solution (\ref{eq:sol1}). All the remaining equations in (\ref{eq:cond2}) to be satisfied
now read as
\begin{align}
\frac{\del c\:\;\;}{\del b_{1}^{(0)}}+2\frac{b_{1}^{+}c+d_{0}}{(b_{1}^{+})^{2}}
 =\frac{\del c\:\;\;}{\del b_{0}^{(0)}}=0,
\end{align}
and their general solution is given by
\begin{align}
b_{1}^{+}c=c_{0}-d_{0}\ln|b_{1}^{+}|,
\label{eq:sol2}
\end{align}
where $c_{0}$ is a constant which does not depend on $b_{1}^{(0)}$. Substituting
(\ref{eq:sol1}) and (\ref{eq:sol2}) back into (\ref{eq:para2}), we have
\begin{align}
d_{1}=a_{2},\qquad a_{1}=c_{0}-d_{0}\ln|b_{1}^{+}|,\qquad a_{0}=0,\qquad
 \bar{c}=d_{0}/b_{1}^{+}.
\label{eq:sol2'}
\end{align}
With the aid of the solutions (\ref{eq:sol1}) and (\ref{eq:sol2'}), we finally obtain the
admissible form of the function $A(z)$ and the parameters $b_{i}^{(2)}$ and $\tR_{2}$ as
\begin{align}
\begin{split}
&A(z)=a_{2}z^{2}+c_{0}z+d_{0}z\ln|z|,\quad b_{1}^{(2)}=b_{1}^{(0)}+2a_{2},\\
&b_{0}^{(2)}=-b_{0}^{(0)}=d_{0},\quad \tR_{2}=-b_{1}^{(0)}-a_{2}.
\label{eq:adm2}
\end{split}
\end{align}
The pair of potentials (\ref{eq:V+-}) in this case are given by
\begin{align}
\lefteqn{
16V^{\pm}(x;b_{1},b_{0})A(z(x))=4\bigl[(a_{2})^{2}+(b_{1})^{2}\bigr] z^{2}
 +4(a_{2}c_{0}+2a_{2}d_{0}\mp 2c_{0}b_{1}\pm 2d_{0}b_{1}}\hspace{30pt}\notag\\
&\;\pm 4a_{2}b_{0}+2b_{1}b_{0})z+3(c_{0})^{2}+2c_{0}d_{0}
 +3(d_{0})^{2}\pm 8(c_{0}+d_{0})b_{0}+4(b_{0})^{2}\notag\\
&\;+4d_{0}(a_{2}\mp 2b_{1})z\ln|z|+2d_{0}(3c_{0}+d_{0}\pm 4b_{0})\ln|z|
 +3(d_{0})^{2}(\ln|z|)^{2}\bigr|_{z=z(x)}.
\label{eq:C22V}
\end{align}
We can easily check that the latter pair satisfies
\begin{align}
V^{+}(x;b_{1},b_{0})A(z)=&\;V^{-}(x;b_{1}+2a_{2},-b_{0})A(z)\notag\\
&\;-(b_{1}+a_{2})\left[a_{2}z^{2}+(c_{0}-d_{0}-b_{0})z+d_{0}z\ln|z|\right],
\end{align}
and thus has two-step SI if and only if $b_{0}=-d_{0}$, as has been indicated
by the third equality in (\ref{eq:adm2}). Hence, the present system has conditional
two-step SI.

We note that $b_{1}\neq0$ in this case and the system always admits two intermediate
Hamiltonians. The first and second intermediate potentials (\ref{eq:Vi1}) and
(\ref{eq:Vi2}) are respectively calculated as
\begin{align}
\lefteqn{
16V^{\rmi1}(x;b_{1},b_{0})A(z(x))=4\bigl[(a_{2})^{2}+(b_{1})^{2}\bigr] z^{2}
 +4(a_{2}c_{0}+2b_{1}b_{0})z-(c_{0}-d_{0})^{2}}\hspace{30pt}\notag\\
&\;+4(b_{0})^{2}+4a_{2}d_{0}z\ln|z|-2d_{0}(c_{0}-d_{0})\ln|z|-(d_{0})^{2}(\ln|z|)^{2}\bigr|_{z=z(x)},
\end{align}
and
\begin{align}
V^{\rmi2}(x;b_{1},b_{0})=&\;V^{\rmi1}(x;b_{1},b_{0})+\frac{(c_{0}-d_{0})b_{1}-2a_{2}b_{0}
 +d_{0}b_{1}\ln|z|}{b_{1}z+b_{0}}\notag\\\
&\;-2b_{0}\frac{c_{0}b_{1}-a_{2}b_{0}+d_{0}b_{1}\ln|z|}{(b_{1}z+b_{0})^{2}}\biggr|_{z=z(x)}.
\end{align}
Both of them do not have ordinary SI with $V^{\pm}$ given in (\ref{eq:C22V})
irrespective of whether $b_{0}=-d_{0}$ or not.
Hence, the conditional two-step SI of the system (\ref{eq:C22V}) is irreducible.
Unfortunately, the equation (\ref{eq:zq}) with $A(z)$ given in (\ref{eq:adm2}) cannot
be integrated analytically in general for $d_{0}\neq0$.\\

In the next, let us consider the case when $b_{1}^{(0)}$ and $b_{0}^{(0)}$ are dependent.
In this case, the general solutions to (\ref{eq:cond1'}) are given by (\ref{eq:cond1''})
with (\ref{eq:dsol1}). However, with the latter solutions $\bar{c}$ will be zero by
the last equality in (\ref{eq:para2}), and we have a quadratic $A(z)$ as a result.
Hence, the system with the dependent $b_{1}^{(0)}$ and $b_{0}^{(0)}$ just reduces to
a system in Case 1-3.\\

\noindent
\textbf{Case 2-3:} $\mu=2$

In this case, we obtain from (\ref{eq:Az2})
\begin{align}
A(z)=a_{2}z^{2}+a_{1}z+a_{0}+\bar{c}(b_{1}^{+}z+b_{0}^{+})^{2}\ln|b_{1}^{+}z+b_{0}^{+}|,
\end{align}
where
\begin{align}
\begin{split}
&a_{2}=c(b_{1}^{+})^{2},\quad a_{1}=2cb_{1}^{+}b_{0}^{+}-\frac{b_{1}^{+}b_{0}^{-}-b_{1}^{-}
 b_{0}^{+}}{2b_{1}^{+}},\\
&a_{0}=c(b_{0}^{+})^{2}-\frac{b_{1}^{+}b_{0}^{-}-b_{1}^{-}b_{0}^{+}}{2(b_{1}^{+})^{2}}b_{0}^{+},
 \quad \bar{c}=\frac{b_{1}^{-}}{2(b_{1}^{+})^{2}}.
\end{split}
\label{eq:para3}
\end{align}
and $c$ is an integral constant. The condition (\ref{eq:cAz2}) in this case is
equivalent to the following set of equations:
\begin{align}
&\frac{\del a_{2}\:\:}{\del b_{i}^{(0)}}+\bar{c}b_{1}^{+}\frac{\del b_{1}^{+}\:\:}{\del b_{i}^{(0)}}
 =\frac{\del a_{1}\:\:}{\del b_{i}^{(0)}}+\bar{c}b_{1}^{+}\frac{\del b_{0}^{+}\:\:}{\del b_{i}^{(0)}}
 +\bar{c}b_{0}^{+}\frac{\del b_{1}^{+}\:\:}{\del b_{i}^{(0)}}=\frac{\del a_{0}\:\:}{\del b_{i}^{(0)}}
 +\bar{c}b_{0}^{+}\frac{\del b_{0}^{+}\:\:}{\del b_{i}^{(0)}}\notag\\
&=b_{1}^{+}\frac{\del \bar{c}\:\;\;}{\del b_{i}^{(0)}}+2\bar{c}\frac{\del b_{1}^{+}\:\:}{\del
 b_{i}^{(0)}}=b_{0}^{+}\frac{\del \bar{c}\:\;\;}{\del b_{i}^{(0)}}
 +2\bar{c}\frac{\del b_{0}^{+}\:\:}{\del b_{i}^{(0)}}=0.
\label{eq:cond3}
\end{align}
We assume $\bar{c}\neq0$ to avoid duplication of Case 1-3.
Substituting the expression of the parameters (\ref{eq:para3}) into (\ref{eq:cond3}),
we see that the condition (\ref{eq:cond1'}) must hold also in the present case. So, let
us first investigate the case when $b_{1}^{(0)}$ and $b_{0}^{(0)}$ are independent. In
this case, by following the same argument below (\ref{eq:cond1'}), we arrive at the same
solution (\ref{eq:sol1}). All the remaining equations in (\ref{eq:cond2}) to be satisfied
now read as
\begin{align}
\frac{\del c\:\;\;}{\del b_{1}^{(0)}}+2\frac{2(b_{1}^{+})^{2}c+d_{1}}{(b_{1}^{+})^{3}}
 =\frac{\del c\:\;\;}{\del b_{0}^{(0)}}=0,
\end{align}
and their general solution is given by
\begin{align}
(b_{1}^{+})^{2}c=c_{0}-d_{1}\ln|b_{1}^{+}|,
\label{eq:sol3}
\end{align}
where $c_{0}$ is a constant which does not depend on $b_{1}^{(0)}$. Substituting
(\ref{eq:sol1}) and (\ref{eq:sol3}) back into (\ref{eq:para3}), we have
\begin{align}
a_{2}=c_{0}-d_{1}\ln|b_{1}^{+}|,\qquad a_{1}=-d_{0},\qquad a_{0}=0,\qquad
 \bar{c}=d_{1}/(b_{1}^{+})^{2}.
\label{eq:sol3'}
\end{align}
With the aid of the solutions (\ref{eq:sol1}) and (\ref{eq:sol3'}), we finally obtain the
admissible form of the function $A(z)$ and the parameters $b_{i}^{(2)}$ and $\tR_{2}$ as
\begin{align}
\begin{split}
&A(z)=c_{0}z^{2}+a_{1}z+d_{1}z^{2}\ln|z|,\qquad b_{1}^{(2)}=b_{1}^{(0)}+2d_{1},\\
&b_{0}^{(2)}=-b_{0}^{(0)}=-a_{1},\qquad \tR_{2}=0.
\label{eq:adm3}
\end{split}
\end{align}
The pair of potentials (\ref{eq:V+-}) in this case are given by
\begin{align}
\lefteqn{
16V^{\pm}(x;b_{1})A(z(x))=\left[4(c_{0})^{2}+3(d_{1})^{2}\pm 8d_{1}b_{1}+4(b_{1})^{2}
 \right]z^{2}+2[a_{1}(2c_{0}-3d_{1})}\hspace{30pt}\notag\\
&\;\mp 4a_{1}b_{1}\pm4(2c_{0}+d_{1})b_{0}+4b_{1}b_{0}]z+3(a_{1})^{2}\pm 8a_{1}b_{0}
 +4(b_{0})^{2}\notag\\
&\;+8c_{0}d_{1}z^{2}\ln|z|+4d_{1}(a_{1}\pm 4b_{0})z\ln|z|+4(d_{1})^{2}z^{2}(\ln|z|)^{2}
 \bigr|_{z=z(x)}.
\label{eq:C23V}
\end{align}
We can easily check that the latter pair satisfies
\begin{align}
V^{+}(x;b_{1},b_{0})A(z)=V^{-}(x;b_{1}+2d_{1},-b_{0})A(z)+(b_{0}-a_{1})(b_{1}+d_{1})z,
\end{align}
and thus has two-step SI if and only if $b_{0}=a_{1}$, as has been indicated
by the third equality in (\ref{eq:adm3}). Hence, the present system has conditional
two-step SI.

We note that $b_{1}\neq0$ in this case and the system always admits two intermediate
Hamiltonians. The first and second intermediate potentials (\ref{eq:Vi1}) and
(\ref{eq:Vi2}) are respectively calculated as
\begin{align}
\lefteqn{
16V^{\rmi1}(x;b_{1},b_{0})A(z(x))=\bigl[4(c_{0})^{2}+8c_{0}d_{1}-(d_{1})^{2}+4(b_{1})^{2}
 \bigr] z^{2}}\hspace{30pt}\notag\\
&\;+2(2a_{1}c_{0}+5a_{1}d_{1}+4b_{1}b_{0})z-(a_{1})^{2}+4(b_{0})^{2}\notag\\
&\;+8d_{1}(c_{0}+d_{1})z^{2}\ln|z|+4a_{1}d_{1}z\ln|z|+4(d_{1})^{2}z^{2}(\ln|z|)^{2}\bigr|_{z=z(x)},
\label{eq:C23i1}
\end{align}
and
\begin{align}
V^{\rmi2}(x;b_{1},b_{0})=&\;V^{\rmi1}(x;b_{1},b_{0})-d_{0}+\frac{a_{1}b_{1}
 -(2c_{0}-d_{1})b_{0}-2d_{1}b_{0}\ln|z|}{b_{1}z+b_{0}}\notag\\\
&\;-2b_{0}\frac{a_{1}b_{1}-c_{0}b_{0}-d_{1}b_{0}\ln|z|}{(b_{1}z+b_{0})^{2}}\biggr|_{z=z(x)}.
\label{eq:C23i2}
\end{align}
Both of them do not have ordinary SI with $V^{\pm}$ given in (\ref{eq:C23V})
irrespective of whether $b_{0}=a_{1}$ or not.
Hence, the conditional two-step SI of the system (\ref{eq:C23V}) is irreducible.

The equation (\ref{eq:zq}) with $A(z)$ given in (\ref{eq:adm3}) cannot be integrated
analytically unless $a_{1}=0$. When $a_{1}=0$, the system automatically has two-step
SI since $b_{0}^{(2)}=b_{0}^{(0)}=0$ from the third equality in (\ref{eq:adm3}). In
addition, the form of $A(z)$ reduces to the one given in (\ref{eq:adm3'}) with $a_{1}=0$,
and the solution to (\ref{eq:zq}) is given by (\ref{eq:C23z}).
Substituting (\ref{eq:C23z}) into (\ref{eq:C23V}), (\ref{eq:C23i1}), and (\ref{eq:C23i2}),
we finally obtain the potentials in the $x$-space.\\

In the next, let us consider the case when $b_{1}^{(0)}$ and $b_{0}^{(0)}$ are dependent.
In this case, the general solutions to (\ref{eq:cond1'}) are given by (\ref{eq:cond1''})
with (\ref{eq:dsol1}). All the remaining equations in (\ref{eq:cond1}) are satisfied with
the solution (\ref{eq:sol3}). Substituting (\ref{eq:cond1''}), (\ref{eq:dsol1}), and
(\ref{eq:sol3}) back into (\ref{eq:para3}), we find that $a_{2}\neq0$ is necessary for
the non-triviality of the system and obtain
\begin{align}
a_{2}=c_{0}-d_{1}\ln|b_{1}^{+}|,\qquad 2a_{2}z_{1}=a_{1},\qquad 4a_{2}a_{0}=(a_{1})^{2}.
\label{eq:dsol3}
\end{align}
With the aid of the solutions (\ref{eq:cond1''}), (\ref{eq:dsol1}), and (\ref{eq:dsol3}),
we finally obtain the admissible form of the function $A(z)$ and the parameters
$b_{1}^{(2)}$ and $\tR_{2}$ as
\begin{align}
\begin{split}
&A(z)=\left(z+\frac{a_{1}}{2a_{2}}\right)^{2}\left(c_{0}+d_{1}\ln\Bigl|z+\frac{a_{1}}{2a_{2}}
 \Bigr|\right),\qquad b_{1}^{(2)}=b_{1}^{(0)}+2d_{1},\\
&b_{0}^{(2)}=\frac{a_{1}}{2a_{2}}b_{1}^{(2)},\qquad\qquad b_{0}^{(0)}=\frac{a_{1}}{2a_{2}}
 b_{1}^{(0)},\qquad\qquad \tR_{2}=0.
\end{split}
\label{eq:adm3'}
\end{align}
The pair of potentials (\ref{eq:V+-}) in this case with the definition $V(x;b_{1}):=
V(x;b_{1},a_{1}b_{1}/(2a_{2}))$ are calculated as
\begin{align}
16V^{\pm}(x;b_{1})A(z(x))=&\;\left(z+\frac{a_{1}}{2a_{2}}\right)^{2}\Biggl[4(c_{0})^{2}
 +(d_{1}\pm 2b_{1})(3d_{1}\pm 2b_{1})\notag\\
&\;+8c_{0}d_{1}\ln\Bigl|z+\frac{a_{1}}{2a_{2}}\Bigr|+4(d_{1})^{2}\left(\ln\Bigl|
 z+\frac{a_{1}}{2a_{2}}\Bigr|\right)^{2}\Biggr]\Biggr|_{z=z(x)}.
\label{eq:C23V'}
\end{align}
We can easily check that the latter pair satisfies
\begin{align}
V^{+}(x;b_{1})=V^{-}(x;b_{1}+2d_{1}),
\end{align}
which is consistent with the formulas for $b_{1}^{(2)}$ and $\tR_{2}$ in (\ref{eq:adm3'}),
and thus has two-step SI. The first intermediate potential (\ref{eq:Vi1}) is given by
\begin{align}
16V^{\rmi1}(x;b_{1})A(z)=&\;\left(z+\frac{a_{1}}{2a_{2}}\right)^{2}\Biggl[4(c_{0})^{2}
 +8c_{0}d_{1}-(d_{1})^{2}+4(b_{1})^{2}\notag\\
&\;+8(c_{0}+d_{1})d_{1}\ln\Bigl|z+\frac{a_{1}}{2a_{2}}\Bigr|+4(d_{1})^{2}\left(\ln\Bigl|
 z+\frac{a_{1}}{2a_{2}}\Bigr|\right)^{2}\Biggr]\Biggr|_{z=z(x)}.
\label{eq:C23i1'}
\end{align}
Intriguingly, the second intermediate potential (\ref{eq:Vi2}) which always exist thanks 
to $b_{1}\neq0$ differs from the first one only by a constant, $V^{\rmi2}(x;b_{1})=
V^{\rmi1}(x;b_{1})-d_{1}$. Both of them do not have ordinary SI with $V^{\pm}$ given
in (\ref{eq:C23V'}). Hence, the present system has irreducible two-step SI.

The equation (\ref{eq:zq}) with $A(z)$ given in (\ref{eq:adm3'}) is integrated as
\begin{align}
\ln\Bigl|z+\frac{a_{1}}{2a_{2}}\Bigr|=\frac{d_{1}}{2}x^{2}-\frac{c_{0}}{d_{1}}.
\label{eq:C23z}
\end{align}
Substituting (\ref{eq:C23z}) into (\ref{eq:C23V'}) and (\ref{eq:C23i1'}), we finally obtain
the potentials in the $x$-space.\\

\noindent
\textbf{Case 3:} $b_{1}^{+}=0$

In this case, we obtain from (\ref{eq:Az2})
\begin{align}
2A(z)=\rme^{\nu z}\int\rmd z\,(-2b_{1}^{(0)}z+b_{0}^{-})\,\rme^{-\nu z},
\end{align}
where $\nu=2\bar{R}/b_{0}^{+}$. We can assume without any loss of generality that
$\nu\neq0$. Indeed, if $\nu=0$, we have
\begin{align}
2A(z)=-b_{1}^{(0)}z^{2}+b_{0}^{-}z+2c,
\end{align}
where $c$ is an integral constant, and thus it just reproduces the case of quadratic $A(z)$.
When $\nu\neq0$, we obtain
\begin{align}
A(z)=a_{1}z+a_{0}+c\,\rme^{\nu z},
\end{align}
where $c$ is an integral constant and
\begin{align}
a_{1}=\frac{b_{1}^{(0)}}{\nu},\qquad a_{0}=\frac{2b_{1}^{(0)}-b_{0}^{-}\nu}{2\nu^{2}}.
\label{eq:para4}
\end{align}
The condition (\ref{eq:cAz2}) in this case is equivalent to the following set of equations:
\begin{align}
&\frac{\del a_{1}\:\:}{\del b_{i}^{(0)}}=\frac{\del a_{0}\:\:}{\del b_{i}^{(0)}}
 =\frac{\del c\:\;\;}{\del b_{i}^{(0)}}=\frac{\del \nu\:\;\;}{\del b_{i}^{(0)}}=0.
\label{eq:cond4}
\end{align}
The last two equalities just mean that both $c$ and $\nu$ do not depend on $b_{i}^{(0)}$.
We assume $c\neq0$ to avoid duplication of Case 1-2.
Substituting the expression of the parameters (\ref{eq:para4}) into the first two equalities
in (\ref{eq:cond4}), we have
\begin{align}
\frac{\del b_{1}^{(0)}}{\del b_{i}^{(0)}}=0,\qquad \frac{\del b_{0}^{-}\:\:}{\del b_{i}^{(0)}}=0.
\label{eq:cond4'}
\end{align}
The first equation in (\ref{eq:cond4'}) is inconsistent if $b_{1}^{(0)}$ is an independent
variable. Thus, the system has only one independent variable $b_{0}^{(0)}$, and $b_{1}^{(0)}$
is a function of it. Then, we immediately have $b_{1}^{(0)}=d_{1}$ where $d_{1}$ is
a constant which does not depend on $b_{0}^{(0)}$. The same equality must hold
also for $b_{1}^{(2)}$ for the consistency, $b_{1}^{(2)}=d_{1}$. On the other hand,
$b_{1}^{+}=b_{1}^{(2)}+b_{1}^{(0)}=0$ in the present case and thus $b_{1}^{(2)}=b_{1}^{(0)}=0$
is the only permissible solution.
The general solution to the second in (\ref{eq:cond4'}) is obviously $b_{0}^{(2)}=b_{0}^{(0)}
+2d_{0}$ where $d_{0}$ does not depend on $b_{0}^{(0)}$. Substituting these solutions back
into (\ref{eq:para4}), we obtain
\begin{align}
a_{1}=0,\qquad d_{0}=-a_{0}\nu.
\end{align}
With the aid of the latter solutions, we finally obtain the admissible form of the function
$A(z)$ and the parameters $b_{i}^{(2)}$ and $\tR_{2}$ as
\begin{align}
A(z)=a_{0}+c\,\rme^{\nu z},\qquad b_{0}^{(2)}=b_{0}^{(0)}-2a_{0}\nu,\qquad
 \tR_{2}=(b_{0}^{(0)}-a_{0}\nu)\nu.
\label{eq:adm4}
\end{align}
The pair of potentials (\ref{eq:V+-}) in this case are given by
\begin{align}
16V^{\pm}(x;b_{0})A(z(x))=-c^{2}\nu^{2}\rme^{2\nu z}-4c\nu(a_{0}\nu\mp2b_{0})\,
 \rme^{\nu z}+4(b_{0})^{2}\bigr|_{z=z(x)}.
\label{eq:C3V}
\end{align}
We can easily check that they satisfy the two-step SI condition (\ref{eq:tssi}) with
the parameters $b_{0}^{(2)}$ and $\tR_{2}$ given in (\ref{eq:adm4}). In this case, $b_{1}=0$
and the system admits only one intermediate Hamiltonian. The intermediate potential
(\ref{eq:Vi1}) is calculated as
\begin{align}
16V^{\rmi1}(x;b_{0})A(z(x))=3c^{2}\nu^{2}\rme^{2\nu z}+4a_{0}c\nu^{2}\rme^{\nu z}
 +4(b_{0})^{2}.
\label{eq:C3i1}
\end{align}
Although it has the same functional dependence on the variable $z$ as $V^{\pm}$
but is not ordinary SI. Hence, the system is irreducible two-step SI.

The equation (\ref{eq:zq}) with $A(z)$ given in (\ref{eq:adm4}) is integrated as
\begin{align}
c\,\rme^{\nu z}=\left\{
\begin{array}{ll}
 \displaystyle{a_{0}\left(\tanh\sqrt{\frac{a_{0}}{2}}\nu x-1\right)}\qquad
  &\text{for}\qquad a_{0}\neq 0,\\
 \displaystyle{\frac{2}{\nu^{2}x^{2}}}&\text{for}\qquad a_{0}=0.
\end{array}\right.
\label{eq:C3z}
\end{align}
Substituting (\ref{eq:C3z}) into (\ref{eq:C3V}) and (\ref{eq:C3i1}), we finally obtain
the potentials in the $x$-space.

\section{Discussion and Summary}
\label{sec:discus}

In this paper, we have studied and constructed two-step SI potentials by utilizing
the framework of $\cN$-fold SUSY. Recognizing the crucial fact that two-step SI
always means type A $2$-fold SUSY with an intermediate Hamiltonian, we have
successfully revealed its general aspects and made the systematic construction of
such systems. The essential point of our analysis resides in the fact that type A
$\cN$-fold SUSY systems possess the simple general form (\ref{eq:V+-}) in terms
of the variable $z$. It enables us to make the model-independent analysis
without recourse to any specific assumption or ansatz. Furthermore, we have
obtained even such two-step SI systems that admit an analytical expression only
in terms of $z$, namely, in Case 2-1 with $a_{1}\neq0$, Case 2-2, and Case 2-3 with
$a_{1}\neq0$. In Table~\ref{tb:1}, we summarize the inequivalent two-step SI
systems and their properties obtained in this paper.

\begin{table}
\begin{center}
\tabcolsep=10pt
\begin{tabular}{ll}
%\hline
% & \\
\hline
Case 1-1 & Reducible\\
Case 1-2 & Reducible\\
Case 1-3 & Reducible\\
Case 1-4 & Irreducible, Conditional ($b_{1}^{-}=-2a_{2}$)\\
             & Irreducible ($b_{1}^{-}\neq -2a_{2}$)\\
Case 1-5 & Irreducible, Conditional ($a_{3}=0$)\\
             & Reducible ($a_{3}\neq 0$)\\
Case 2-1 & Irreducible, Conditional (2 independent parameters)\\
             & Irreducible (1 independent parameter, $\mu\neq 4$)\\
Case 2-2 & Irreducible, Conditional\\
Case 2-3 & Irreducible, Conditional (2 independent parameters)\\
             & Irreducible (1 independent parameter)\\
Case 3 & Irreducible\\
\hline
\end{tabular}
\caption{List of the obtained two-step SI potentials. Cases 1-4 and 1-5 can be
included in Case 2-1 ($\mu=3$ and $\mu=4$, respectively).}
\label{tb:1}
\end{center}
\end{table}

It is remarkable that all the obtained two-step SI as well as ordinary SI potentials
are of translational classes, that is, parameter relations are characterized by constant
shifts $b_{i}^{(2)}=b_{i}^{(0)}+2d_{i}$. In the traditional approaches, one assumes
the latter relation from the beginning and then tries to construct an SI potential
which meets it. A significant point of our analysis is that it is a consequence of
the differential equation $\del b_{i}^{-}/\del b_{i}^{(0)}=0$ originated from the
requirement (\ref{eq:cAz2}) that $A(z)$ should not depend on the parameters
$b_{i}^{(0)}$. It in particular means that another class of SI potentials which is
different from translational ones could be obtainable only from such a $A(z)$ that
depends explicitly on $b_{i}^{(0)}$. Then, to obtain a scaling SI potential which has
parameter relations $b_{i}^{(2)}=q_{i}b_{i}^{(0)}$, for instance,
SI conditions are expected to lead to a differential equation like
\begin{align*}
\frac{\del\:\:}{\del b_{i}^{(0)}}(\ln b_{i}^{(2)}-\ln b_{i}^{(0)})=0.
\end{align*}
In this way, we would be able to make a systematic analysis also on non-translational SI
potentials.

In \cite{BDGKPS93}, the concept of multi-step SI was also introduced as a natural
extension of two-step SI. Following a similar argument in Section~\ref{sec:tssi}, we easily
conclude that multi-step SI is a special case of $\cN$-fold SUSY with intermediate
Hamiltonians at every intermediate positions for $\cN>2$. Until now on, there has
been only one systematic study on such systems in the case of type A $3$-fold
SUSY~\cite{BT10}.
It was shown in the latter reference that every type A $3$-fold SUSY systems with
intermediate Hamiltonians at every intermediate positions, which were dubbed Class (1,1),
were ordinary SI as well. That is, any three-step SI in type A $3$-fold SUSY is reducible.
Hence, irreducible three-step SI can be realized, if it exists, only in other types of $3$-fold
SUSY.

Another interesting aspect of our present results which can be immediately read from
Table~\ref{tb:1} is that all the conditional two-step SI are irreducible. To the best of our
knowledge, there have been no decisive mathematical understanding of conditional
exact solvability. The result thus indicates the possibility that the latter concept might
be well characterized in a context of irreducible two- and multi-step SI.

Although we have restricted our investigation to ordinary scalar Schr\"{o}dinger
operators, the concepts of ordinary, two-step, and multi-step SI are applicable to
much wider systems.
In fact, $\cN$-fold SUSY was successfully formulated also for Schr\"{o}dinger operators
with position-dependent mass~\cite{Ta06a}, matrix ones~\cite{Ta11b}, and ones with
reflection operators~\cite{Ta11c}. Hence, we would be able to make systematic studies on
various SI in these systems with the framework of $\cN$-fold SUSY, as have been done
in this work.

%\section*{Acknowledgments}
%\begin{acknowledgments}%REVTEX4
%\ack%IOPART

%\end{acknowledgments}%REVTEX4

%\appendix

%\section*{References}%IOPART

\bibliography{refsels}%BIB-FILE

\begin{thebibliography}{10}
\expandafter\ifx\csname url\endcsname\relax
  \def\url#1{{\tt #1}}\fi
\expandafter\ifx\csname urlprefix\endcsname\relax\def\urlprefix{URL }\fi
\providecommand{\eprint}[2][]{\url{#2}}

\bibitem{Wi82}
E.~Witten, Nucl. Phys. B 202 (1982) 253.

\bibitem{CF83}
F.~Cooper and B.~Freedman, Ann. Phys. 146 (1983) 262.

\bibitem{Ge83}
L.~{\'{E}}. Gendenshte{\^{\i}}n, JETP Lett. 38 (1983) 356.

\bibitem{IH51}
L.~Infeld and T.~E. Hull, Rev. Mod. Phys. 23 (1951) 21.

\bibitem{GK85}
L.~{\'{E}}. Gendenshte{\u{\i}}n and I.~V. Krive, Sov. Phys. Usp. 28 (1985) 645.

\bibitem{WAG89}
J.~Wu, Y.~Alhassid, and F.~G{\"{u}}rsey, Ann. Phys. 196 (1989) 163.

\bibitem{GMS98}
A.~Gangopadhyaya, J.~V. Mallow, and U.~P. Sukhatme, Phys. Rev. A 58 (1998)
  4287.

\bibitem{CDGPRS98}
S.~Chaturvedi, R.~Dutt, A.~Gangopadhyaya, P.~Panigrahi, C.~Rasinariu, and
  U.~Sukhatme, Phys. Lett. A 248 (1998) 109.
\newblock {E}rratum-ibid. 251 (1999) 406, \eprint{arXiv:hep-th/9807081}.

\bibitem{Ba98}
A.~B. Balantekin, Phys. Rev. A 57 (1998) 4188.
\newblock \eprint{arXiv:quant-ph/9712018}.

\bibitem{CGK87}
F.~Cooper, J.~N. Ginocchio, and A.~Khare, Phys. Rev. D 36 (1987) 2458.

\bibitem{DKS88}
R.~Dutt, A.~Khare, and U.~P. Sukhatme, Am. J. Phys. 56 (1988) 163.

\bibitem{Le89}
G.~L{\'{e}}vai, J. Phys. A: Math. Gen. 22 (1989) 689.

\bibitem{Ch91}
C.~Chuan, J. Phys. A: Math. Gen. 24 (1991) L1165.

\bibitem{BDGKPS93}
D.~T. Barclay, R.~Dutt, A.~Gangopadhyaya, A.~Khare, A.~Pagnamenta, and
  U.~Sukhatme, Phys. Rev. A 48 (1993) 219.
\newblock \eprint{arXiv:hep-ph/9304313}.

\bibitem{KS93}
A.~Khare and U.~P. Sukhatme, J. Phys. A: Math. Gen. 26 (1993) L901.
\newblock \eprint{arXiv:hep-th/9212147}.

\bibitem{BGM10}
J.~Bougie, A.~Gangopadhyaya, and J.~Mallow, Phys. Rev. Lett. 105 (2010) 210402.
\newblock \eprint{arXiv:1008.2035 [hep-th]}.

\bibitem{DK02}
M.~Daoud and M.~Kibler, Phys. Part. Nucl. 33 (2002) S43.

\bibitem{CF03}
A.~Chenaghlou and H.~Fakhri, Int. J. Mod. Phys. A 18 (2003) 939.
\newblock \eprint{arXiv:hep-th/0203240}.

\bibitem{DK04}
M.~Daoud and M.~Kibler, Phys. Lett. A 321 (2004) 147.
\newblock \eprint{arXiv:math-ph/0312019}.

\bibitem{DK06}
M.~Daoud and M.~Kibler, J. Math. Phys. 47 (2006) 122108.
\newblock \eprint{arXiv:quant-ph/0609017}.

\bibitem{GMS01}
A.~Gangopadhyaya, J.~V. Mallow, and U.~P. Sukhatme, Phys. Lett. A 283 (2001)
  279.
\newblock \eprint{arXiv:hep-th/0103054}.

\bibitem{MR09b}
B.~Midya and B.~Roy, J. Phys. A: Math. Theor. 42 (2009) 285301.
\newblock \eprint{arXiv:0910.3179 [quant-ph]}.

\bibitem{Su08a}
W.-C. Su, J. Phys. A: Math. Theor. 41 (2008) 255307.

\bibitem{Su08b}
W.-C. Su, J. Phys. A: Math. Theor. 41 (2008) 435301.

\bibitem{Su09}
W.-C. Su, J. Phys. A: Math. Theor. 42 (2009) 385202.

\bibitem{AIS93}
A.~A. Andrianov, M.~V. Ioffe, and V.~P. Spiridonov, Phys. Lett. A 174 (1993)
  273.
\newblock \eprint{arXiv:hep-th/9303005}.

\bibitem{AST01b}
H.~Aoyama, M.~Sato, and T.~Tanaka, Nucl. Phys. B 619 (2001) 105.
\newblock \eprint{arXiv:quant-ph/0106037}.

\bibitem{AS03}
A.~A. Andrianov and A.~V. Sokolov, Nucl. Phys. B 660 (2003) 25.
\newblock \eprint{arXiv:hep-th/0301062}.

\bibitem{TU87}
A.~V. Turbiner and A.~G. Ushveridze, Phys. Lett. A 126 (1987) 181.

\bibitem{Us94}
A.~G. Ushveridze, {Q}uasi-exactly {S}olvable {M}odels in {Q}uantum {M}echanics
  (IOP Publishing, Bristol, 1994).

\bibitem{AST01a}
H.~Aoyama, M.~Sato, and T.~Tanaka, Phys. Lett. B 503 (2001) 423.
\newblock \eprint{arXiv:quant-ph/0012065}.

\bibitem{GT06}
A.~Gonz{\'a}lez-L{\'o}pez and T.~Tanaka, J. Phys. A: Math. Gen. 39 (2006) 3715.
\newblock \eprint{arXiv:quant-ph/0602177}.

\bibitem{BB98a}
C.~M. Bender and S.~Boettcher, Phys. Rev. Lett. 80 (1998) 5243.
\newblock \eprint{physics/9712001}.

\bibitem{Mo02a}
A.~Mostafazadeh, J. Math. Phys. 43 (2002) 205.
\newblock \eprint{arXiv:math-ph/0107001}.

\bibitem{Ta06a}
T.~Tanaka, J. Phys. A: Math. Gen. 39 (2006) 219.
\newblock \eprint{arXiv:quant-ph/0509132}.

\bibitem{Ta09}
T.~Tanaka, In Morris~B. Levy, ed., Mathematical Physics Research Developments
  (Nova Science Publishers, Inc., New York, 2009), chapter~18. pp. 621--679.

\bibitem{BT09}
B.~Bagchi and T.~Tanaka, Ann. Phys. 324 (2009) 2438.
\newblock \eprint{arXiv:0905.4330 [hep-th]}.

\bibitem{BT10}
B.~Bagchi and T.~Tanaka, Ann. Phys. 325 (2010) 1679.
\newblock \eprint{arXiv:1002.1766 [hep-th]}.

\bibitem{RS88}
V.~A. Rubakov and V.~P. Spiridonov, Mod. Phys. Lett. A 3 (1988) 1337.

\bibitem{SoDu93}
A.~de~Souza~Dutra, Phys. Rev. A 47 (1993) R2435.

\bibitem{Ta11b}
T.~Tanaka.
\newblock {$\mathcal{N}$}-fold supersymmetry in quantum mechanical matrix
  models.
\newblock To appear in {\textit{Mod. Phys. Lett.}}, \eprint{arXiv:1108.0480
  [math-ph]}.

\bibitem{Ta11c}
T.~Tanaka.
\newblock {$\mathcal{N}$}-fold supersymmetric quantum mechanics with
  reflections.
\newblock \eprint{arXiv:1112.0087 [math-ph]}.

\end{thebibliography}
\bibliographystyle{npb}%BST-FILE
%\begin{thebibliography}{99}

%\def\J#1#2#3#4{{\sl #1} {\bf #2} (#3) #4}

%\bibitem{}
%Author 1, Author 2 and Author 3,
%\J{Journal}{Volume}{Year}{Page}.

%\end{thebibliography}

\end{document}